\documentclass[final,3p,times]{elsarticle}

\usepackage{amssymb}

\usepackage{graphicx}
\usepackage{amsmath}
\usepackage{subfigure}
\usepackage{subfigmat}
\usepackage{natbib}

 \def\vector#1{\mbox{\boldmath $#1$}}
 
 \def\tensor#1{\mathcal #1}

\journal{Journal of Computational Physics}

\makeatletter
\def\@author#1{\g@addto@macro\elsauthors{\normalsize%
    \def\baselinestretch{1}%
    \upshape\authorsep#1\unskip\textsuperscript{%
      \ifx\@fnmark\@empty\else\unskip\sep\@fnmark\let\sep=,\fi
      \ifx\@corref\@empty\else\unskip\sep\@corref\let\sep=,\fi
      }%
    \def\authorsep{\unskip,\space}%
    \global\let\@fnmark\@empty
    \global\let\@corref\@empty  
    \global\let\sep\@empty}%
    \@eadauthor={#1}
}
\makeatother

\makeatletter
\def\ps@pprintTitle{%
   \let\@oddhead\@empty
   \let\@evenhead\@empty
   \let\@oddfoot\@empty
   \let\@evenfoot\@oddfoot
}

\begin{document}

\begin{frontmatter}


\title{Eulerian-Lagrangian method for simulation of cloud cavitation}
\author{Kazuki Maeda\corref{cor1}}
\ead{maeda@caltech.edu}
\cortext[cor1]{Corresponding author}

\author{Tim Colonius}



\address{Division of Engineering and Applied Science, California Institute of Technology\\
1200 East California Boulevard, Pasadena, CA 91125, USA}

\begin{abstract}
We present a coupled Eulerian-Lagrangian method to simulate
cloud cavitation in a compressible liquid.
The method is designed to capture the strong, volumetric oscillations of each bubble and the bubble-scattered acoustics.
The dynamics of the bubbly mixture is formulated using volume-averaged equations of motion.
The continuous phase is discretized on an Eulerian grid and integrated using a high-order, finite-volume weighted essentially non-oscillatory (WENO) scheme, while the gas phase is modeled as spherical, Lagrangian point-bubbles at the sub-grid scale, each of whose radial evolution is tracked by solving the Keller-Miksis equation.
The volume of bubbles is mapped onto the Eulerian grid as the void fraction by using a regularization (smearing) kernel.
In the most general case, where the bubble distribution is arbitrary, three-dimensional Cartesian grids are used for spatial discretization.
In order to reduce the computational cost for problems possessing translational or rotational homogeneities, we spatially average the governing equations along the direction of symmetry and discretize the continuous phase on two-dimensional or axi-symmetric grids, respectively.
We specify a regularization kernel that maps the three-dimensional distribution of bubbles onto the field of an averaged two-dimensional or axi-symmetric void fraction.
A closure is developed to model the pressure fluctuations at the sub-grid scale as synthetic noise.
For the examples considered here,  modeling the sub-grid pressure fluctuations as white noise agrees a priori with computed distributions from three-dimensional simulations, and suffices, a posteriori, to accurately reproduce the statistics of the bubble dynamics.
The numerical method and its verification are described by considering test cases of the dynamics of a single bubble and cloud cavitaiton induced by ultrasound fields.
\end{abstract}


\end{frontmatter}


\section{Introduction}
\label{S:1}
Cavitation and bubble cloud dynamics are of importance in various
fields of engineering, including therapeutic
ultrasound\citep{Coleman87,Pishchalnikov03,Ikeda06}, hydraulic
machineries\citep{Naude61,Brennen13}, and underwater acoustics\citep{Etter13}:
cavitation bubbles formed in a human body during a passage of the tensile part of ultrasound pulses can scatter and absorb subsequent pulses to lower the efficiency of ultrasound therapies and violent collapse of bubbles can cause cavitation damage on the surface of various materials, from ship propellers to kidney stones.
Bubbles that naturally exist in the ocean can attenuate and modulate acoustic signals to decrement the performance of underwater acoustic systems.
Accurate simulations of cloud cavitation are in high-demand for such applications, yet they are challenging due to the complex, multi-scale nature of the interactions among the dynamics of small, dispersed bubbles and pressure waves propagating in the liquid.
Numerical methods that fully resolve the bubble-liquid interface (which we denote as ``direct methods''), includes interface-capturing methods with high-order/interface-sharpening schemes\citep{Johnsen06,Shukla10,Tiwari13,Coralic14,Shyue14}, level set/ghost fluid\citep{Fedkiw99,HU06}, front tracking\citep{Terashima09} and Mixed-Eulerian-Lagrangian/boundary integral methods\citep{Calvisi07,Wang10}. These methods can in principle accurately represent the full dynamics, but their high computational cost of resolving the interfaces, such methods are the most suitable for simulations at the scales of bubbles for a short period of time, typically single events of the collapse of bubbles \citep{Lauer12,Rossinelli13,Tiwari15,Rasthofer17}.

For more complex bubbly mixtures with relatively low void fractions of $O(10^{-2})$, methods that solve volume- or ensemble- averaged equations of motion have been used to compute the propagation of acoustic and shock waves\citep{Biesheuvel84,Caflisch85,Commander89,Zhang94,Kameda96,Colonius00,Smereka02,Seo10,Ando11,Grandjean12} and the dynamics of bubble clouds \citep{Omta87,dAgostino89,Wang99,Tanguay02,Matsumoto05,Fuster11,Ma18}.
In the classical averaging approaches, the bubbly-mixture is treated as a continuous media.
The volume of dispersed bubbles is converted to a continuous void fraction field in a control volume that contains a sufficiently large number of bubbles. The length-scale of the control volume (averaging length-scale) is larger than the characteristic inter-bubble distance.
The dynamics of the gas-phase are closed by considering the averaged change in the volumetric oscillations of bubbles in response to pressure fluctuations in the mixture.
Bubbles are typically modeled as spherical cavities, of which dynamics are described by ordinary differential equations (ODEs) in a form of the Rayleigh-Plesset (RP) equation\citep{Plesset77}.
The radius and coordinate of the bubbles are treated as statistically averaged quantities in the field of mixture, rather than deterministic variables defined at each single bubble.
Inter-bubble interactions can be modeled by an {\it{effective}} pressure in the mixture that forces the oscillations of bubbles, enabling the methods to avoid explicitly solving for the interactions.
Meanwhile, due to the construction of the model, the field of the mixture smaller than the averaging length-scale is unresolved.
This limitation hinders accurate simulations of cloud cavitation induced by a pressure wave with a wavelength as small as characteristic inter-bubble distance.
Such cavitation bubble cloud is observed in experiments using high-intensity focused ultrasound and particularly important for applications to ultrasound therapies\citep{Maeda15}.
Even though the incident wave can be resolved, bubble-scattered pressure waves generated by violent collapses can have finer scales.
Such waves can be likewise unresolved, despite their importance for assessment of cavitation damage on nearby surfaces.

Aside from the averaging methods, Lagrangian point-bubble approaches can directly solve for the dynamics of interacting bubbles, by extending the Rayleigh-Plesset equation for a system of multiple bubbles in an incompressible or weakly compressible potential flow under far-field pressure excitation\citep{Takahira94,Doinikov04,Bremond06,Zeravcic11,Thomas12}.
The methods treat bubbles as Lagrangian points whose coordinate and radius are explicitly tracked, and thus can capture the entire scales of the flow field of the surrounding liquid at the scale of bubbles.
Meanwhile, the methods are, in general, valid with far-field pressure uniform at the coordinates of the entire bubbles.
This requirement for the scale separation limits applications of the methods to bubble clouds excited by pressure waves with a wavelength much greater than the size of the cloud.

A coupled Eulerian-Lagrangian method has been recently developed to solve for the dynamics of the bubbly-mixture in a wide range of scales\citep{Fuster11}.
The scheme combines the averaging approach and the Lagrangian approach; it solves the dynamics of the mixture on an Eulerian grid, while tracks the dynamics of Lagrangian point bubbles at the sub-grid scale.
Notably, the method is free from the spatial limitation in the averaging approaches as well as from the constraint on the wavelength of the far-field pressure in the Lagrangian point-bubble approaches.
By using a grid with a high resolution, the method can resolve the incident wave and the bubble-scattered pressure waves on the grid at a scale smaller than the inter-bubble distance, while accurately track the radial evolutions of the bubbles.
Yet, the method employed numerical schemes suitable for resolving weak pressure disturbances in a linear (acoustic) regime.
As a result, the applications of the methods were limited in relatively mild oscillations of bubbles that result in a small amplitude of bubble-scattered pressure waves that the schemes can support.
Moreover, the method was applied to simulations of bubble clouds with two-dimensional geometries including cylinder and ellipse, and three-dimensional clouds with realistic geometries were not explored.

The goal of the present study is to generalize and extend the Eulerian-Lagrangian method for simulation of cloud cavitation induced by pressure waves with a strong amplitude and a high-frequency for practical applications.
In the method, the dynamics of bubbly-mixture is described using the volume-averaged equations of motion that fully account for the compressibility of liquid.
The continuous phase is discretized on an Eulerian grid, while
the gas phase is modeled as spherical, radially oscillating cavities that are tracked as Lagrangian points at the sub-grid scale.
The dynamics of the continuous phase is evolved using a high-order, finite-volume weighted essentially non-oscillatory (WENO) scheme, that was originally developed for simulation of viscous,
compressible, multi-component flows\citep{Coralic14} and is capable of capturing a strong pressure waves with fine structures.
The volume of bubbles is mapped onto the Eulerian grids as the void fraction using a regularization kernel.
The radial oscillation of each bubble is evolved by solving the Keller-Miksis equation.
When the the grid size is smaller than the characteristic inter-bubble distance, the method is capable of capturing the violent cavitation growth and collapse of each bubble as well as resolving the strong, complex structures of bubble-scattered pressure waves in the liquid.

For the most general cases, the continuous phase is discretized on a three-dimensional Cartesian grid, and a standard regularization kernel is used to map the volume of bubbles onto the field of void fraction.
To reduce the cost for simulations of a bubbly-mixture that possesses translational or axi-symmetric homogeneity, we newly introduce reduced-order models.
In the models, the continuous phase is discretized on two-dimensional or axi-symmetric grids. We map the volume of bubbles distributed in three-dimensional space
onto the two-dimensional or axi-symmetric field of void fraction by using a modified regularization kernel.
By doing so, the cost of computations required to solve for the continuous phase is reduced from $O(N^3)$ to $O(N^2)$, where $N$ is the number of grid cell per dimension, in comparisons to the three-dimensional model.

In order to properly close the Keller-Miksis equation, the pressure
field at the sub-grid scale needs to be appropriately modeled.
In the case of the three-dimensional model, in each grid cell that encloses a bubble, the contribution of the pressure wave scattered by the bubble to the averaged pressure in the cell can become significant, thus the pressure of the cell cannot be directly used to force the oscillations of bubble. In that case, following the scheme proposed by Fuster and Colonius\citep{Fuster11}, we obtain the component of the cell-averaged pressure that forces the oscillations of bubble by using the state of the bubble and potential flow theory at the sub-grid scale.
In the two-dimensional and axi-symmetric models, the discretized pressure field is treated as uniform in the direction of symmetry, despite the three-dimensionality of the true pressure field associated with any distribution of bubbles.
In order to reduce the error associated with the neglected three-dimensional pressure fluctuations, we model the spatial distribution of the pressure at the sub-grid scale as white noise.
In each grid cell that contains a bubble, we estimate the variance of the noise by sampling the pressure in the neighboring cells, with an assumption that the pressure fluctuations are locally, spatially isotropic on the scale of the sampling window. We express the noise by superposing Fourier modes with pre-computed, randomized phases, following a method of expressing stochastic fluctuations in turbulence modeling\citep{Bechara94,Smirnov01}.
The sub-grid closures for the three-dimensional and the reduced models are verified using the test cases of acoustic cavitation of a single bubble and a bubble screen, and a bubble cloud, respectively.

Finally, the methods are used to simulate a challenging case of cloud cavitation excited in a strong ultrasound wave. The structure of the bubble cloud obtained in the simulation is confirmed to qualitatively agree with a bubble cloud observed in high-speed images, that was excited by a focused ultrasound generated by a medical transducer.
The paper proceeds as follows. In Sec. \ref{section:gov_eqn}, we introduce volume-averaged equations of motion. Then we describe the discretization and spatial integration of the governing equations on a three-dimensional Cartesian grid and regularization of the volume of bubbles using a kernel. Subsequently we describe the dynamical equations of the radial oscillations of bubbles as a closure.
In Sec. \ref{section:reduce}, we describe the reduction of the governing equations on two-dimensional and axi-symmetric coordinates and introduce regularization kernels for the grids. In Sec. \ref{section:result}, we present numerical tests of single bubble and cloud cavitation for verifications of the methods. Finally, in Sec. \ref{section:conc}, we state a summary and conclusions along with suggestions for future work.

\section{Governing equations}
\label{section:gov_eqn}
\subsection{Volume averaged equations of motion}
We introduce volume-averaged equations of motion to describe the dynamics of a mixture of dispersed bubbles and a compressible liquid in three-dimensional space.
Volume-averaged equations consider the conservation of mass, momentum and energy of the mixture as a continuum media that are defined by applying the volume averaging operator $\overline{(\cdot)}$ to a control volume of the mixture:
$\overline{(\cdot)}=(1-\beta)(\cdot)_l+\beta(\cdot)_g$,
where $\beta\in[0,1)$ is the volume fraction of gas (void fraction), and
subscripts $l$ and $g$ denote the liquid and gas phase, respectively.
We start by writing the equations in a conservative form:
\begin{align}
\frac{\partial \overline\rho}{\partial
t}+\nabla\cdot(\overline{\rho\vector{u}})
&=0,\label{eqn:vaeqn_ma}\\
\frac{\partial (\overline{\rho\vector{u}})}{\partial t} + \nabla\cdot
(\overline{\rho\vector{u}}\otimes\overline{\vector{u}}+p\tensor{I}
-\tensor{T})
&=0,\label{eqn:vaeqn_mo}\\
\frac{\partial \overline{E}}{\partial t} + \nabla\cdot
\left((\overline{E}+p)\overline{\vector{u}}
-\tensor{T}\cdot\overline{\vector{u}}\right)
&=0\label{eqn:vaeqn_e},
\end{align}
where $\rho$ is the density, $\vector{u}=(u,v,w)^{\mathrm{T}}$ is the velocity,
$p$ is the pressure and $E$ is the total energy, respectively.
$\tensor{T}$ is the effective viscous stress tensor of the mixture.
We invoke two approximations widely used in averaged models at the limit of low void fraction, up to $O(10^{-2})$\citep{Caflisch85,Commander89,Fuster11}.
First, the density of liquid is typically much larger than that of gas, $\rho_l \gg \rho_g$, thus the density of the mixture is approximated by that of the liquid:
\begin{equation}
\overline\rho=(1-\beta)\rho_l+\beta\rho_g\approx(1-\beta)\rho_l\label{eqn:rho_app}.
\end{equation}
This approximation is clearly valid for the mixture of water and air/vapor bubbles under practical conditions.
Second, the slip velocity between the two phases is zero:
\begin{equation}
\overline{\vector{u}}\approx\vector{u}_l=\vector{u}_g\label{eqn:u_app}.
\end{equation}
With the assumption of zero slip-velocity, the momentum flux across the gas-liquid interface is effectively zero, and therefore, we approximate the total viscous stress as that in the continuous phase:
\begin{equation}
\tensor{T}=(1-\beta)\tensor{T}_l\label{eqn:tau_app}.
\end{equation}

$\tensor{T}_l$ is the viscous stress tensor of pure Newtonian liquid:
\begin{equation}
\tensor{T}_l=2\mu\left(\tensor{D}_l
-\frac{1}{3}(\nabla\cdot\vector{u}_l)\tensor{I}\right),
\end{equation}

where $\mu_l$ is the shear viscosity of liquid and $\tensor{D}_l$ is the
deformation rate tensor:

\begin{equation}
\tensor{D}_l=\frac{1}{2}(\vector{u}_l+\vector{u}_l^{\mathrm{T}}).
\end{equation}

We note that, in reality, spherical bubbles experience hydrodynamic forces from the surrounding liquid\citep{Magnaudet98}, and the resulting slip velocity can be non-zero.
The momentum flux across the gas-liquid interface can contribute to
the effective viscosity of the mixture\citep{Zhang94}.
Such modeling is not a focus of the present study, though one could extend the
present formulation to include the effect of the non-zero slip velocity on
$\tensor{T}$.
Nevertheless, for many practical problems of cavitation, the time scale of the radial oscillations of bubbles are estimated to be much shorter than that of the translational motions, and therefore, assumption of the zero-slip velocity is a reasonable first approximation\citep{Caflisch85}.

Using relations (\ref{eqn:rho_app}-\ref{eqn:tau_app}), equations
(\ref{eqn:vaeqn_ma}-\ref{eqn:vaeqn_e}) can be rewritten
as conservation equations in terms of the mass, momentum and energy of the liquid
with source terms, as an inhomogeneous hyperbolic system:
\begin{align}
\frac{\partial \rho_l}{\partial t} + \nabla\cdot (\rho_l
\vector{u}_l)
&=\frac{\rho_l}{1-\beta}\left[\frac{\partial
\beta}{\partial t}+\vector{u}_l\cdot\nabla\beta\right],\label{eqn:vaeqn2_ma}\\
\frac{\partial (\rho_l\vector{u}_l)}{\partial t} +
\nabla\cdot (\rho_l\vector{u}_l\otimes \vector{u}_l+p\tensor{I}-\tensor{T}_l)
&=\frac{\rho_l\vector{u}}{1-\beta}\left[\frac{\partial
\beta}{\partial
t}+\vector{u}_l\cdot\nabla\beta\right]
-\frac{\beta\nabla\cdot(p\tensor{I}-\tensor{T}_l)}{1-\beta},\label{eqn:vaeqn2_mo}\\
\frac{\partial E_l}{\partial t} + \nabla\cdot
\left((E_l+p)\vector{u}_l-\tensor{T}_l\cdot\vector{u}_l\right)
&=\frac{E_l}{1-\beta}\left[\frac{\partial
\beta}{\partial
t}+\vector{u}_l\cdot\nabla\beta\right]
-\frac{\beta\nabla\cdot(p\vector{u}_l-\tensor{T}_l\cdot\vector{u}_l)}{1-\beta}.
\label{eqn:vaeqn2_e}
\end{align}
This form of equations is particularly convenient since we can directly apply the finite volume WENO scheme
for spatial integration of the equations, which will be discussed in the following section.
For later convenience, we also denote the equations in a vector form:
\begin{equation}
\frac{\partial \vector{q}_l}{\partial t} + \nabla\cdot
\vector{f}(\vector
q_l)=\vector{g}(\vector{q_l},\beta,\dot{\beta})\label{eqn:vaeqn_vec},
\end{equation} 
where $\vector{q}_l=[\rho_l,\rho_l\vector{u}_l,E_l]$ and
\begin{equation}
\vector{g}=\frac{1}{1-\beta}\frac{d\beta}{dt}\vector{q}_l
-\frac{\beta}{1-\beta}\nabla\cdot(\vector{f}-\vector{u}_l\vector{q}_l).
\end{equation}
For a thermodynamic closure for the liquid, we employ stiffened gas equation of state:
\begin{equation}
p=(\gamma-1)\rho\varepsilon-\gamma\pi_\infty,
\label{eqn:stiff_gas}
\end{equation}
where $\varepsilon$ is the internal energy of liquid, $\gamma$ is the specific
heat ratio, and $\pi_\infty$ is the stiffness, respectively.
In the present study we use $(\gamma , \pi_\infty) = (7.1, 3.06 × 10^8)$ for water.
At the limit of small change in the density of liquid, the equation of state can
be linearized as
\begin{equation}
p=p_0+c_0^2(\rho-\rho_0)
\label{eqn:lin_eqs},
\end{equation}
where
\begin{equation}
c=\sqrt{\gamma(p+\pi_\infty)/\rho}
\label{eqn:c}
\end{equation}
is the speed of sound in liquid and
the subscript 0 denotes reference states.

\subsection{Spatial discretization}
In the following we describe a method of numerical representation of the
governing equation in $x-y-z$ 3D Cartesian coordinate.
We spatially discretize equation (\ref{eqn:vaeqn_vec}):

\begin{equation}
\frac{\partial \vector{q}_l}{\partial t}
+\frac{\partial \vector{f}^x(\vector{q}_l)}{\partial x}
+\frac{\partial \vector{f}^y(\vector{q}_l)}{\partial y}
+\frac{\partial \vector{f}^z(\vector{q}_l)}{\partial z}
=\vector{g},
\end{equation}

where $\vector{f}^x$, $\vector{f}^y$ and $\vector{f}^z$ are vectors of fluxes
in $x$, $y$ and $z$ directions.
We integrate the above equation in arbitrary finite volume grid cell

\begin{equation}
I_{i,j,k}=[x_{i-1/2},x_{i+1/2}]\times[y_{j-1/2},y_{j+1/2}]\times[z_{k-1/2},z_{k+1/2}],
\end{equation}

where $i$, $j$ and $k$ are the indices in $x-$, $y-$ and
$z-$ directions, and $x_{i\pm1/2}$, $y_{j\pm1/2}$ and $z_{k\pm1/2}$ are the
positions of cell faces.
At each finite volume cell, we express the equation in the following
semi-discrete form:

\begin{equation}
\frac{d\vector{q}_{l,i,j,k}}{dt}
=\frac{1}{\Delta x_{i}}[\vector{f}^x_{i-1/2,j,k}-\vector{f}^x_{i+1/2,j,k}]
+\frac{1}{\Delta y_{j}}[\vector{f}^y_{i,j-1/2,k}-\vector{f}^y_{i,j+1/2,k}]
+\frac{1}{\Delta z_{j}}[\vector{f}^z_{i,j,k-1/2}-\vector{f}^z_{i,j,k+1/2}]
+\vector{g}_{i,j,k}.\label{eqn:vae_semi}
\end{equation}

The conservative variables at cell faces are reconstructed by
a high-order WENO scheme from cell-centered values, then are used in HLLC
Riemann solver to calculate the fluxes.
In the present study we employ fifth-order WENO scheme (WENO5) introduced by
Coralic and Colonius \citep{Coralic14}, unless otherwise noted.
High-order WENO schemes are, in general, robust in capturing discontinuities including shock wave and material's interface, while capable of resolving continuous waves with a high-amplitude with relatively small numerical dissipation/dispersion\citep{Titarev04,Pirozzoli06,Shu16}.
Such properties of WENO schemes are suitable for simulations of cloud cavitation in the regime of the interest of the present study;
a passage of strong pressure waves causes violent, nonlinear oscillations of bubbles, each of which emits strong pressure waves with broadband frequency and generate complex structures of pressure fields by mutual interactions.

\subsection{Void fraction}
We express $\vector{g}$, $\beta$ and $\dot{\beta}$ as functions of the state of the bubbles.
To do so, we employ a Lagrangian point-bubble approach, in that the gas phase is modeled as spherical, radially oscillating cavities consisted of a non-condensible gas and liquid vapor.
The center of $n$th bubble ($n\in\mathbb{Z}:n\in[1, N]$), with a radius of
$R_n$ and a radial velocity of $\dot{R}_n$, is initially defined at the coordinate
$\vector{x}_n$ and tracked as Lagrangian points during simulations.
To define the continuous field of the void fraction in the mixture at coordinate
$\vector{x}$, we smear the volume of bubble using a regularization kernel $\delta$:
\begin{equation}
\beta(\vector{x})=\sum^N_{n=1}V_n(R_n)\delta(d_n, h),
\end{equation}
where $V_n$ is the volume of bubble $n$, $V_n=4/3\pi R_n^3$, and $d_n$ is the distance of the coordinate $\vector{x}$ from the center of the bubble, $d_n=|\vector{x}-\vector{x}_n|$.
Various types of kernels are used to regularize Lagrangian variables in particle methods and immersed boundary methods\citep{Cottet00,Peskin02,Monaghan05}, and have also been applied to dispersed bubbly-mixture\citep{Kitagawa01,Fuster11,Ma18}. 
In the present study, we use the continuous, second order, truncated Gaussian function for the kernel:
\begin{equation}
\delta(d_n,h)
=
\left\{
  \begin{array}{@{}ll@{}}
    \frac{1}{h^3(2\pi)^{3/2}}e^{-\frac{d_n^2}{2h^2}},&0\leq\frac{d_n}{h}<3\\
    0&3\leq\frac{d_n}{h}.
  \end{array}\right
  .
\end{equation}

where $h$ is the width of the support of the kernel.
Likewise,
\begin{equation}
\frac{\partial\beta(\vector{x})}{\partial t}
=
\frac{\partial}{\partial t}\sum^N_{n=1}V_n\delta
=
\sum^N_{n=1}\frac{\partial V_n}{\partial
t}\delta+\sum^N_{n=1}V_n\frac{\partial \delta}{\partial
t},
\end{equation}
where
\begin{equation}
\frac{\partial V_p}{\partial
t}=4\pi R^2_n\dot{R}_n,\hspace{10 pt} \frac{\partial \delta}{\partial
t}=-\vector{u}_l\cdot \nabla\delta.
\end{equation}
In the discretized field, we are regularizing the discontinuous distribution of the volume
of a bubble defined at a singular point in a finite volume cell, by distributing the void fraction within neighboring cells around the bubble (fig. \ref{fig:grids}a).

\subsection{Bubble dynamics}
We model the dynamics of volumetric oscillations of the bubbles forced by pressure fluctuations in the surrounding liquid. When the time scale of the evolution of the pressure in the mixture is sufficiently mild and slow, compared to the characteristic frequency of bubbles, the change of the states of the gas inside the bubble can be assumed as being quasi-static.
Meanwhile, in the problems we aim to simulate, bubbles oscillate in response to rapid and high-amplitude changes in the pressure with a high amplitude.
In turn, the bubble oscillations generate and scatter pressure waves into the surrounding liquid.
Moreover, mass-transfer due to phase change as well as heat transfer at the
bubble wall can damp the volumetric oscillations.
To model bubble oscillations, we employ the Keller-Miksis equation, combined with a reduced-order model introduced by Preston et al.\citep{Preston07} for the heat and mass-transfer.
In the model, the states of the vapor-gas mixture are treated uniform in the bubble.
The Keller-Miksis equation is a second order, nonlinear ODE in terms of the radius of a single, isolated bubble in an unbounded, weakly compressible liquid:
\begin{equation}
\left(R_n\left(1-\frac{\dot{R}_n}{c}\right)\right)\ddot{R}_n
+
\frac{3}{2}\dot{R}^2\left(1-\frac{\dot{R}_n}{3c}\right)
=
\frac{p_n-p_\infty}{\rho}
\left(1+\frac{\dot{R}_n}{c}\right)+\frac{R_n\dot{p}_n}{\rho c},\label{eqn:KM}
\end{equation}
\begin{equation}
p_{n}=p_{Bn}-\frac{4\mu_l\dot{R}_n}{R_n}-\frac{2\sigma}{R_n},\label{eqn:pb}
\end{equation}
where $p_n$ is the pressure at the bubble wall,
$p_{Bn}$ is the pressure inside the bubble,
$\sigma$ is the surface tension,
and $p_\infty$ is the component of the pressure that forces the radial oscillations of
the bubble.
The reduced-order model formulates $\dot{p}_n$ and the vapor mass in the bubble $\dot{m}_{Vn}$ as
\begin{align}
\dot{p}_{Bn}
&=\textsf{func}
[R_n,\dot{R}_n,{m}_{Vn}]\label{eqn:preston_pdot}\\
\dot{m}_{Vn}
&=
\textsf{func}[R_n,{m}_{Vn}]\label{eqn:preston_mdot}.
\end{align}
For the explicit forms of equations (\ref{eqn:preston_pdot}) and
(\ref{eqn:preston_mdot}) as well as further details and validation/verification of the model, see Preston et al.\citep{Preston07}.
Overall, equations (\ref{eqn:KM}-\ref{eqn:preston_mdot}) consist a system of
ODEs in terms of $[R_n,\dot{R}_n,p_{Bn},m_{Vn}]$, that can be integrated given initial
conditions and $p_\infty$.
We will discuss the treatment of $p_\infty$ in the next section.
We note that, in principle, other variations of the Rayleigh-Plesset equation as well as
models of heat and mass-transfer could be used to express the sub-grid bubble dynamics.

\subsection{Modeling $p_\infty$: bubble dynamic closure}
We now introduce a closure to model $p_{\infty}$. The Keller-Miksis equation models the pressure field surrounding a single, isolated, spherically symmetric bubble\citep{Keller80}.
$p_\infty$ represents an incoming acoustic wave that drives the oscillations.
The pressure wave scattered by the bubble is also represented by a spherical outgoing wave, $p_{out}$.
The pressure field, as a solution of the Keller-Miksis equation, is thus the superposition of $p_\infty$ and $p_{out}$.
In a bubble cloud, $p_\infty$ for each bubble is the superposition of $p_{out}$ emitted by the surrounding bubbles and the pressure wave that propagate from outside the cloud, and in general not know a-priori.
Following Fuster and Colonius (hereafter denoted as FC)\citep{Fuster11}, we obtain $p_\infty$ for each sub-grid bubble by modeling the pressure field in a finite volume cell that encloses the bubble.
The pressure of the cell (or group of cells) is given by the spatial average of the superposition of $p_\infty$
and $p_{out}$ emitted by the sub-grid bubble:
\begin{equation}
p_{cell}=\frac{1}{V_{l,cell}}\int_{V_{cell}}(p_\infty+p_{out})dv_l\approx
p_\infty+\frac{1}{V_{l,cell}}\int_{V_{cell}}p_{out}dv_l\label{eqn:p_cell}.
\end{equation}
where $V_{l,cell}$ is the volume of the liquid in the control volume $V_{cell}$ and $dv_l$ is the volume element of the liquid, respectively.
A natural choice of $V_{cell}$ is the region of the liquid over which the volume of bubble is smeared over.
\footnote{
For a bubble located in a cell $I_{m,n,p}$, we define that the cells over which the volume of the bubble is smeared as $I_{i,j,k}:i\in[m-N_r,m+N_r],j\in[n-N_r,n+N_r],k\in[p-N_r,p+N_r]$, where $N_r=\lfloor 3h/\Delta \rfloor$.
}
We can assume that
\begin{equation}
\frac{1}{V_{l,cell}}\int_{V_{cell}} p_\infty dv_l\approx p_\infty,
\end{equation}
since $p_\infty$ is approximately uniform at a scale of the cell:
$\Delta\ll\lambda$, where $\lambda$ is the characteristic wave length of the
pressure wave.
Meanwhile, in the discretized field, $p_{cell}$ can be directly approximated as:
\begin{equation}
p_{cell,i,j,k}\approx\frac{\sum^{i+N_r}_{i-N_r}\sum^{j+N_r}_{j-N_r}{\sum^{k+N_r}_{k-N_r}(1-\beta_{m,n,p}){p}_{m,n,p}V_{m,n,p}}}{\sum^{i+N_r}_{i-N_r}\sum^{j+N_r}_{j-N_r}\sum^{k+N_r}_{k-N_r}(1-\beta_{m,n,p})V_{m,n,p}},
\label{eqn:p_dcell}
\end{equation}
where $V_{i,j,k}$ is the volume of cell $I_{i,j,k}$.
$\frac{1}{V_{l,cell}}\int p_{out} dv_l$ can be modeled using the dynamical states of the bubble and $p_{cell}$:
\begin{equation}
\frac{1}{V_{l,cell}}\int_{V_{cell}}p_{out}dv_l\approx\frac{1}{1-C_1}\left[p_{cell}-p_n-\left(C_2-\frac{1}{2}\right)\rho \dot{R}_n^2\right],\label{eqn:p_out}
\end{equation}
where $C_1$ and $C_2$ are functions of $R_n$ and $V_{l,cell}$.
We can spatially discretize this expression to represent $p_{out}$ at each computational cell that
contains a bubble, and substitute into relation (\ref{eqn:p_cell}), along with (\ref{eqn:p_dcell}), to obtain
$p_\infty$. As discussed by FC, $p_{out}$ spatially decay with $r$, where $r$ is the distance from the center of the bubble.
When $V_{cell}$ is much larger than the volume of the
bubble, the contribution of $p_{out}$ to $p_{cell}$ is negligible and
$p_{cell}\approx p_\infty$ holds.
We note the detailed derivation of relation (\ref{eqn:p_out}) in Appendix C.

\subsection{Length scales of parameters}
\label{section_scales}
The present method is designed to correctly capture the small-scale dynamics of cloud cavitation when the following inequalities is satisfied:
\begin{equation}
\left\{
  \begin{array}{@{}ll@{}}
    R_b&\\
    \Delta&
  \end{array}\right\}
\leq h<L_b,\label{ine:scale}
\end{equation}
where $R_b$ is the characteristic bubble radius, $\Delta$ is the characteristic
Eulerian grid size and $L_b$ is the characteristic inter-bubble distance.
With the range of parameters satisfying (\ref{ine:scale}),
we naturally have at most single bubble within each cell.
The inequality between $\beta$ and $h$ comes from the upper bound of $\beta$:
$\beta<1$.
\begin{equation}
\max(\beta_{i,j,k})\sim V_b\max(\delta)
\sim\frac{\max(R_b)^3}{h^3}.
\end{equation}
Therefore
\begin{equation}
\max(\beta_{i,j,k})<1\rightarrow \max(R_b)<h\label{eqn:maxR}.
\end{equation}
Meanwhile, the value of $\max(R_b)$ is not known a-priori, and inequality
 (\ref{eqn:maxR}) may not be not guaranteed to hold for the initially given $h$.
In that case, one may dynamically increase $h$ to satisfy the inequality during the simulation.
The inequality between $\Delta$ and $h$ is a necessary condition for the correct
representation of the regularization kernel on the grid\citep{Cottet00}.
The inequality between $h$ and $L_b$ prevents overlap among the kernel support.
The minimum resolved length scale of waves emitted by the source is $h$;
spatial scales finer than $h$ are smeared by the kernel.
Therefore when the smeared regions of neighboring bubbles overlap with each
other, the pressure field of a scale as small as inter-bubble distance is
likewise filtered.
In that case, as partially discussed by FC, the model tends to recover solutions of classical ensemble
averaged equations, in that the smallest length scale in the field becomes
the wavelength of the pressure waves that propagate in the averaged bubbly-mixture.
The present method is designed to capture the small scales,
and thus the support width is set shorter than the characteristic inter-bubble distance.
Note that $R_b$ can be admissibly larger than $\Delta$ (which means the bubble size can be larger than the grid size) as long as (\ref{eqn:maxR}) is satisfied.

\subsection{Temporal integration}
For temporal integrations of solutions, we employ 4th/5th order
Runge–Kutta-Cash-Karp (RKCK) algorithm\citep{Cash90}.
The stability of the temporal integration of Eulerian variables
is dictated by the Courant-Friedrichs-Lewy (CFL) number, $C$, and the diffusion
number, $D$ (CFL conditions).
The Keller-Miksis equation is a stiff ODE, and in a certain interval of integration it
requires a time step size much smaller than that required by the CFL conditions.
In the problems shown in the present study, we initially set a fixed time step
size that satisfies the CFL condition to satisfy $C<0.3$ and $D<0.15$.
At each time step, we integrate Eulerian variables using the 4th order scheme
built in the algorithm.
Lagrangian variables are updated using the 4th and 5th order
schemes with the same time step size, then the errors between the
two solutions are calculated.
If the errors are smaller than a tolerance, the algorithm employs the 5th
order solution, while if not, both Eulerian/Lagrangian
variables are re-calculated with a smaller value of time step size. We repeat
this process until the error becomes smaller than the tolerance.

\subsection{Acoustic source}
In simulations we excite volumetric oscillations of bubbles using
various amplitudes of traveling pressure waves.
In order to generate the waves, we utilize a source-term approach introduced by Maeda and Colonius\citep{Maeda17}.
The method can generate uni-directional acoustic waves from an arbitrary
geometry of a source-surface, by forcing the mass, momentum
and energy equations (\ref{eqn:vaeqn2_ma}-\ref{eqn:vaeqn2_e})
on the surface in a domain.

\section{Model reduction of the three-dimensional volume-averaged equations}
\label{section:reduce}
In many problems, cloud cavitation occurs in statistically two-dimensional (e.g. flows over a two-dimensional body), or axi-symmetric (e.g. ellipsoidal/spherical bubble cloud) configurations,
in the sense that the flow field and the spatial distribution of bubbles are homogeneous in certain directions.
To simulate the bubbly-mixture in such configurations with lower computational expense, we derive a reduced model by spatially averaging the three-dimensional volume-averaged equations along the direction of symmetry, and then discretize the continuous phase on a two-dimensional/axi-symmetric grid.
In order to properly map the three-dimensional distribution of bubbles onto the void fraction defined in such grid cells, we will introduce modified regularization kernels.
$p_\infty$ is recovered by modeling pressure fluctuations at the sub-grid scale as
locally isotropic, stochastic noise.
The speed-up of simulations achieved by using the reduced models is discussed in Appendix B.

\subsection{Two-dimensional volume averaged equations}
We consider the flow field defined on Cartesian coordinates.
We assume that the flow field and the spatial distribution of bubbles are homogeneous along the $z$-axis.
We define the line-averaging operator $T_{z}$ as
\begin{equation}
T_z(\cdot)=\frac{1}{L}\int^{L/2}_{-L/2}(\cdot)dz,
\end{equation}
where $L$ is the scale of homogeneity.
In order to obtain two-dimensional equations, we apply $T_z$ to
the three-dimensional volume-averaged equations:
\begin{align}
T_z\left[\frac{\partial \vector{q}_l}{\partial t} + \nabla\cdot
\vector{f}(\vector q_l)\right]=T_z\vector{g}(\vector{q}_l,\beta,\dot{\beta}).
\label{eqn:vaeq_symb}
\end{align}
Due to the homogeneity of the flow field along the $z$-axis,
$z$ component of equation (\ref{eqn:vaeq_symb}) is projected onto the nullspace of $T_z$.
$T_z\partial\vector{q}_l/\partial t=\partial(T_z\vector{q}_l)/\partial t$, while $T_z\nabla\cdot\vector{f(\vector{q}_l)}\ne \nabla\cdot\vector{f}(T_z\vector{q}_l)$ and $T_z\vector{g}(\vector{q}_l,\beta,\dot{\beta})\ne \vector{g}(T_z\vector{q}_l,T_z\beta,T_z\dot{\beta})$, since $\vector{f}$ and $\vector{g}$ are nonlinear functions of $\vector{q}_l$, $\beta$ and/or $\dot{\beta}$.
We decompose $T_z\nabla\cdot\vector{f}$ and $T_z\vector{g}$ into linear and nonlinear components:
\begin{align}
T_z\nabla\cdot\vector{f}(\vector{q}_l)
&=\nabla\cdot\vector{f}(T_z\vector{q}_l)+\nabla\cdot\vector{f}_{res},\\
T_z\vector{g}(\vector{q}_l,\beta,\dot{\beta})
&=\vector{g}(T_z\vector{q}_l,T_z\beta,T_z\dot{\beta})+\vector{g}_{res},
\end{align}
where $\vector{f}_{res}$ and $\vector{g}_{res}$ are residuals defined by these equations.
By substituting these expressions, equation (\ref{eqn:vaeq_symb}) becomes
\begin{align}
\frac{\partial (T_z\vector{q}_l)}{\partial t} + \nabla\cdot
\vector{f}(T_z\vector{q}_l)
&=
\vector{g}(T_z\beta,T_z\vector{q_l})-\nabla\cdot\vector{f}_{res}+\vector{g}_{res}\\
&\approx
\vector{g}(T_z\beta,T_z\vector{q_l})+\vector{g}_{res}\label{eqn:vaeq_symb2d},
\end{align}
where we applied $|\nabla\cdot\vector{f}_{res}|\ll|\vector{g}_{res}|$, assuming that the back-ground flow field is uniform at the scale of homogeneity, $L$.
In order to close the equations, $\vector{g}_{res}$ needs to be modeled.
To do so, we decompose $p$ and $\beta$ into its spatial mean and
fluctuation in the $z$ direction:
\begin{align}
p&=T_zp+p'\\
\beta&=T_z\beta+\beta',
\end{align}
where the prime denotes fluctuations. Note that $T_z(\cdot)'=0$.

By substituting these expressions and neglecting terms
higher than 2nd order, $T_z\vector{g}$ can be expressed
as:
\begin{align}
T_z\vector{g}_z(\vector{q}_l,\beta,\dot{\beta})
&=
T_z\left(\frac{1}{1-\beta}\frac{d\beta}{d
t}\vector{q}_l\right)
-T_z\left(\frac{\beta}{1-\beta}\nabla\cdot(\vector{f}-\vector{u}_l\vector{q}_l)\right)\\
&\approx
\vector{g}_z(T_z\vector{q}_l,T_z\beta,T_z\dot{\beta})
+T_z\left(\beta'\frac{d \beta'}{d t}\right)
+(2T_z\beta+1)T_z(\beta'\nabla\cdot(\vector{f}-\vector{u}_l\vector{q}_l)'),
\end{align}
where
$\nabla\cdot(\vector{f}-\vector{u}_l\vector{q}_l)'\approx[0,\nabla
p',(T_z\vector{u}_l)\cdot\nabla p']^{\mathrm{T}}$.
Therefore it is sufficient to model $p'$, $\beta'$ and $\dot{\beta'}$ to express
$\vector{g}_{res}$.

\subsection{Regularization kernel for 2D Cartesian grid}
In order to compute $\beta'=\beta-T_z\beta$
and $\partial\beta'/\partial t=\partial\beta/\partial
t-T_z(\partial\beta'/\partial t)$, we express
\begin{equation}
T_z\beta(x,y)=\frac{1}{L}\int^{L/2}_{-L/2}
\beta(x,y,z) dz
=
\frac{1}{L}\int^{L/2}_{-L/2}
\sum^P_NV_n\delta dz
=
\sum^P_NV_n\left[\frac{1}{L}\int^{L/2}_{-L/2}\delta dz\right]
=
\sum^P_NV_n\delta_{2D},
\end{equation}
\begin{equation}
T_z\frac{d\beta}{dt}=\sum_N^P\frac{d}{dt}\left[V_p\delta_{2D}\right],
\end{equation}
where we defined
\begin{equation}
\delta_{2D}=\frac{1}{L}\int^{L/2}_{-L/2}\delta dz.
\end{equation}
$\delta_{2D}$ can be interpreted as a regularization kernel that maps
the volume of bubbles distributed in three-dimensional space onto
two-dimensional space spanned by $x$-$y$ coordinates basis vectors.

\begin{figure}
 \begin{subfigmatrix}{1}
  \subfigure[]{\includegraphics[width=56mm]{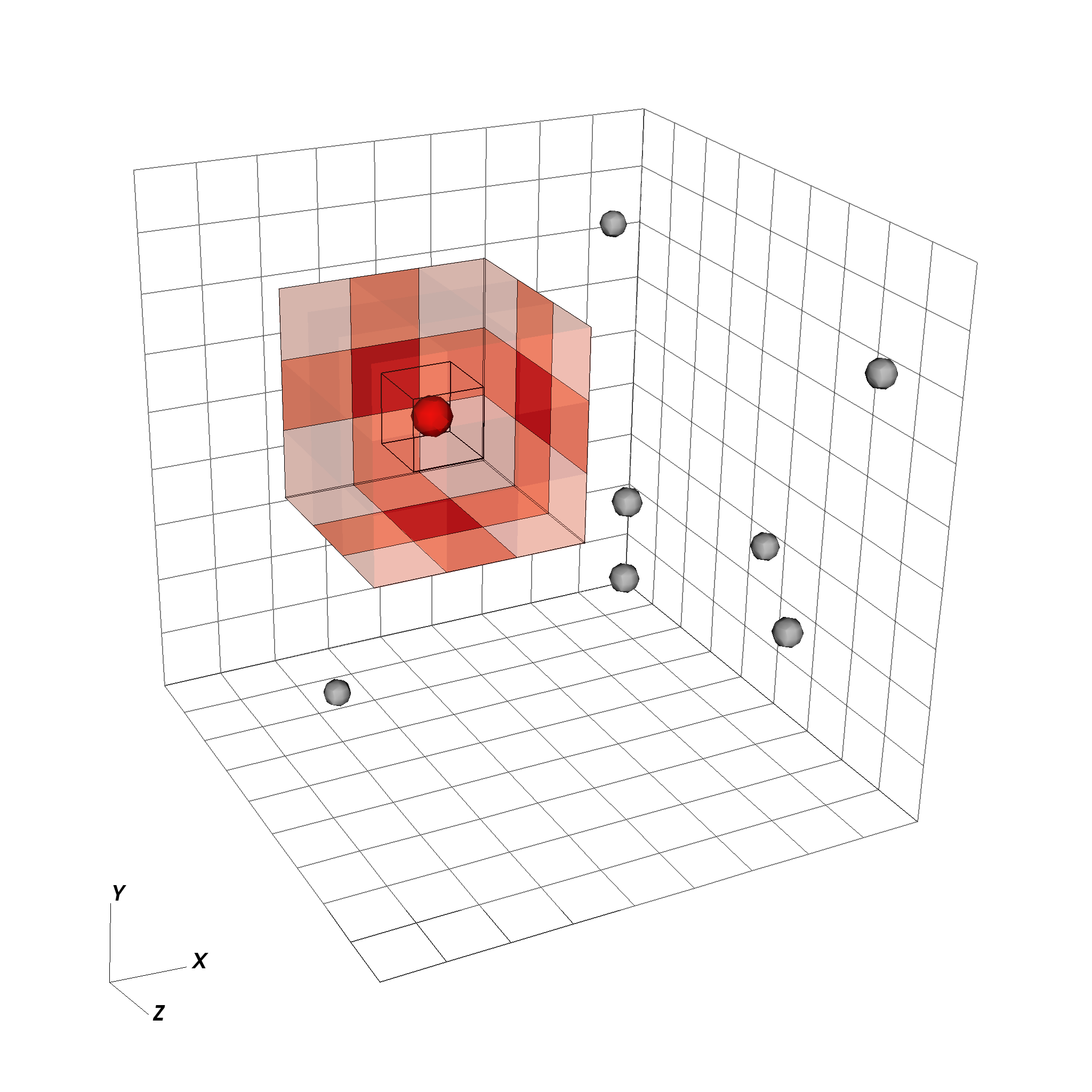}}
  \subfigure[]{\includegraphics[width=56mm]{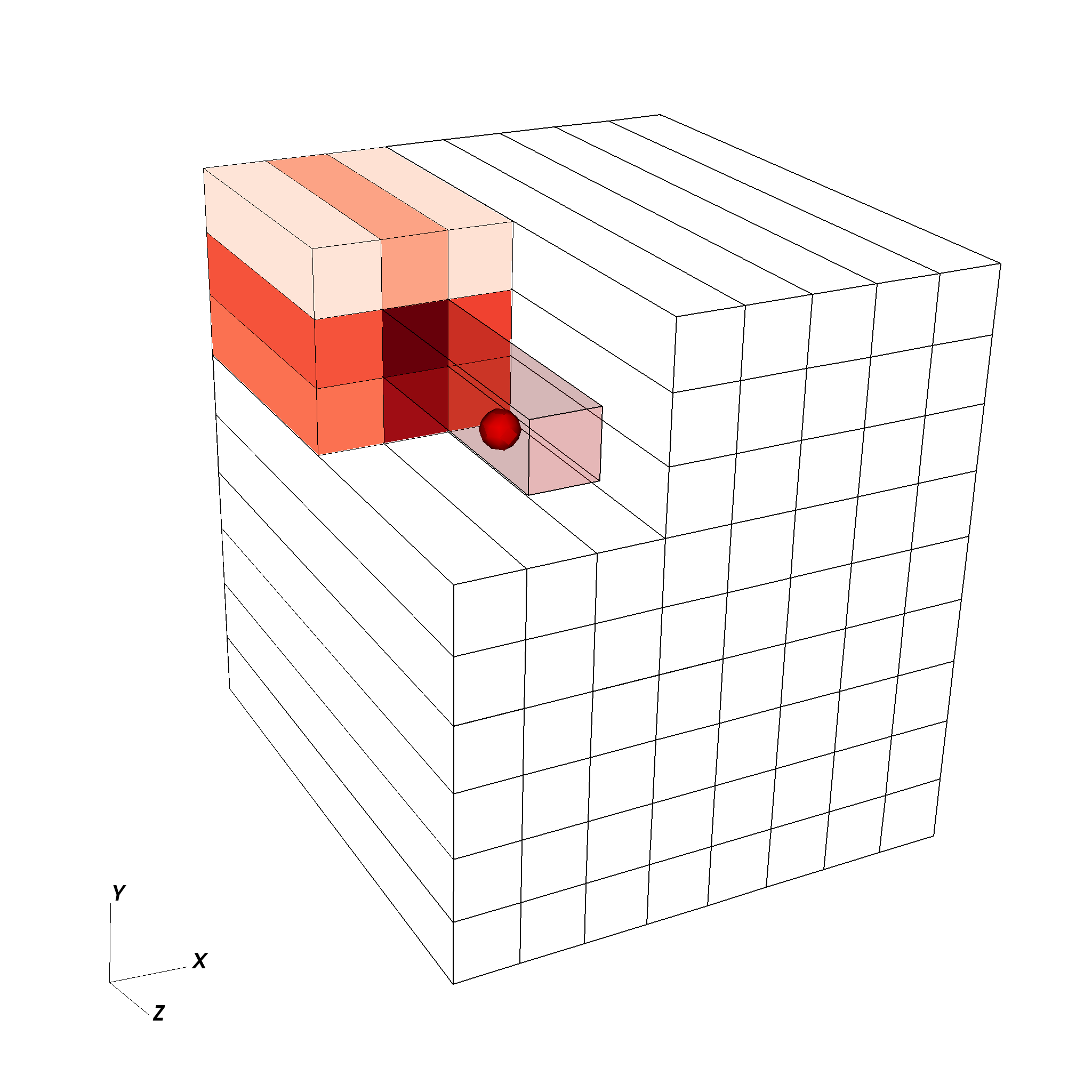}}
  \subfigure[]{\includegraphics[width=50mm]{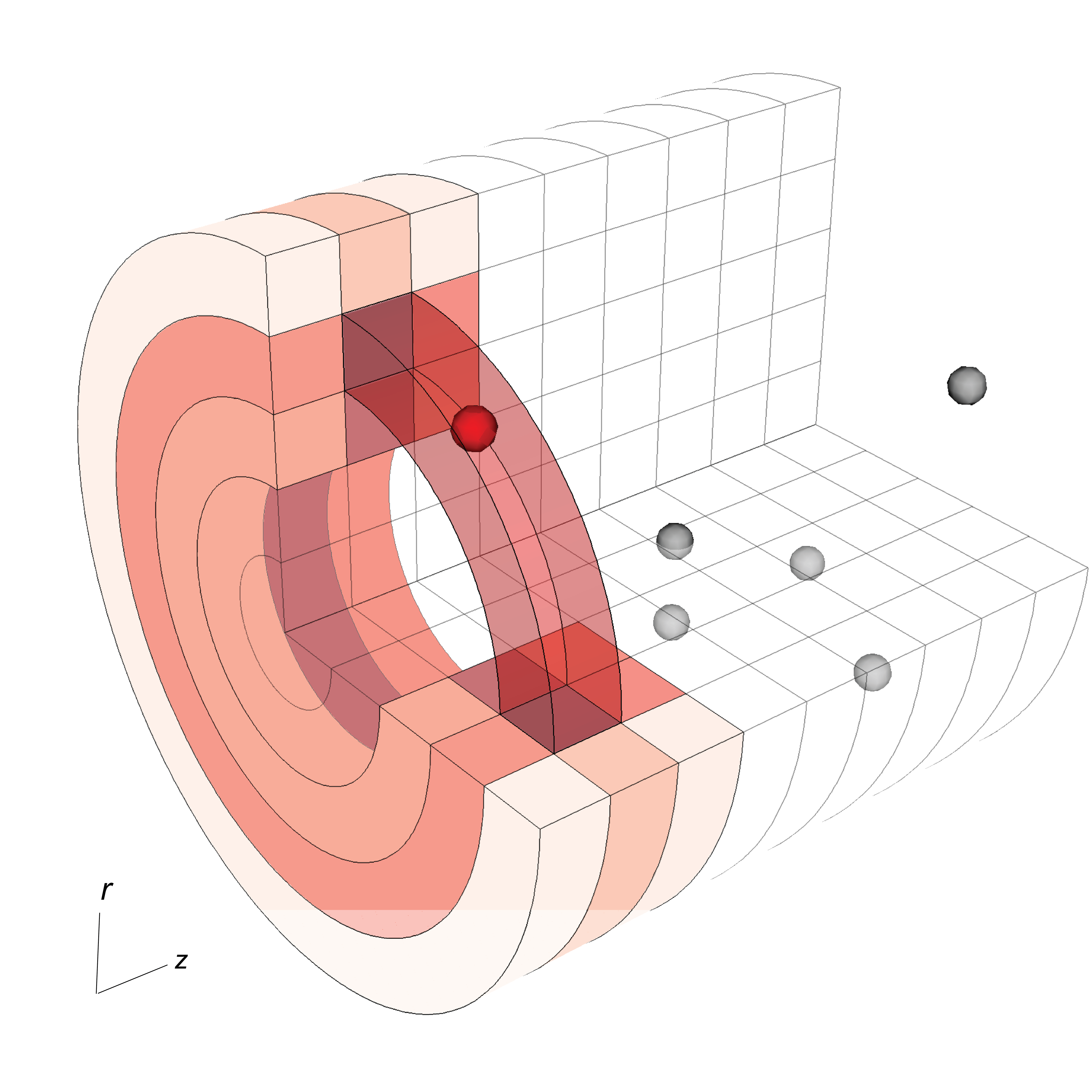}}
  \end{subfigmatrix}
  \caption{Schematic of the smearing of the volume of Lagrangian bubbles on
  neighboring finite volume cells defined on various grids in the same domain:
   (a) three dimensional Cartesian grid;
   (b) two dimensional Cartesian grid;
   (c) axi-symmetric grid.
   On each grid, for the same bubble (red-colored), we are shading the cell that contains the bubble and those neighboring to it, on which the volume of bubble is smeared over as the void fraction. Depth of the shade indicates the value of the void fraction, which decays with the distance from the bubble.
   It is apparent that, depending on the choice of the grid, the volume is mapped onto different regions in the domain.
   }
   \label{fig:grids} 
\end{figure}

In a geometric interpretation, we are essentially solving for the volume averaged
equations with $u_z=0$, on finite volume grid cells in a shape of
parallelepiped with a span of $L$ along the $z$-axis (fig \ref{fig:grids}b).
The Eulerian variables are treated as being uniform in each parallelepiped cell.
However, the physical distributions of the bubbles are three-dimensional and non-uniform in the $z$ direction.
To correct the discrepancy, we are modeling the quadratic, nonlinear terms in terms of
$\beta$ and $\dot{\beta}$, which appear in $\vector{g}_{res}$.
In order to numerically represent $\delta_{2D}$ at each parallelepiped cell,
we discretize the parallelepiped cell into smaller $n_p$ cells in $z$-direction,
and apply a mid-point rule to the integral:
\begin{eqnarray}
\delta_{{2D}_{i,j}}
\approx
\frac{1}{L}\sum_{k=1}^{n_p}\Delta
z_k\delta(d_n(\vector{x}_n,\vector{x}_{i,j,k}),h),
\label{eqn:2dkernel}
\end{eqnarray}
Note that the small cells are essentially identical to those defined on 3D
Cartesian grids, $I_{i,j,k}$.
The total contribution of the volume of particle $P$ on
$\beta_{{2D}_{i,j}}$ is represented by an overlapping region of $I_{i,j}$ and the ball within that $V_n$ is smeared over (fig \ref{fig:grids}b).

\subsection{Modeling $p_\infty$ on the reduced space: stochastic closure}
\begin{figure}
  \center
  \includegraphics[width=100mm,trim = 00mm 0mm
  00mm 0mm,clip]{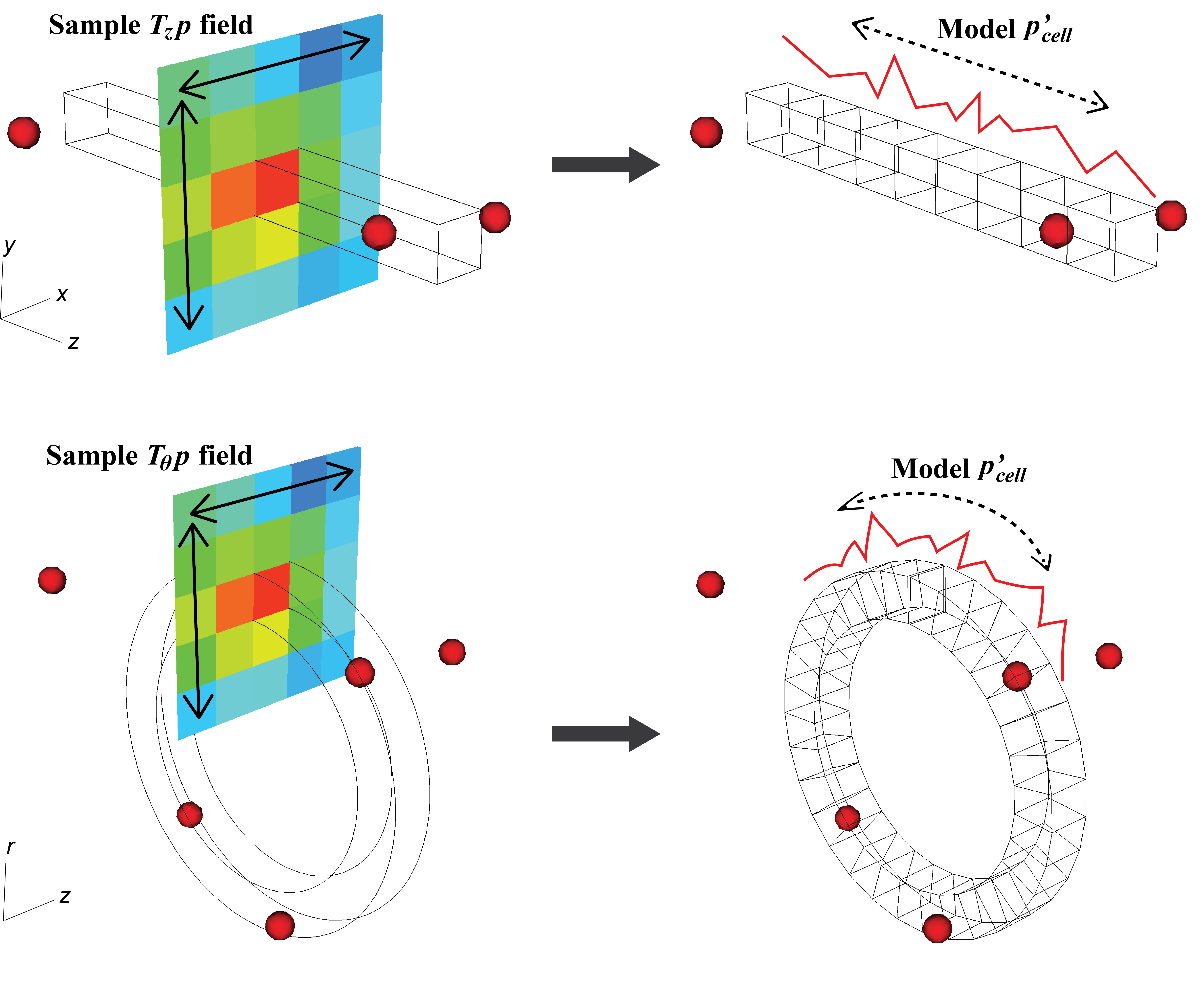}
  \caption{Schematic of the technique to estimate $p'_{cell}$ in the reduced models (top: two-dimensional, bottom: axi-symmetric).}
   \label{fig:sampling} 
\end{figure}
As discussed in section (\ref{section_scales}),
$p_\infty$ needs to be recovered from $p_{cell}=T_zp_{cell}+p_{cell}'$ to
correctly force Lagrangian bubbles.
The first term can be approximated as
$T_zp_{cell}$:
\begin{equation}
T_zp_{cell}=T_z\frac{1}{V_{cell}}\int_{V_{cell}}(p_\infty+p_{out})dv_l\approx
T_zp_\infty+T_z\int_{V_{cell}}p_{out}dv_l\approx T_zp_{\infty}.
\end{equation}
Note that the contribution of $p_{out}$
is negligible compared to $p_\infty$ since $V_{cell}\gg V_n$.
We are missing $p'_{cell}$ and not able to recover it from the
sub-grid dynamics, unlike the closure for the three-dimensional model.
Therefore we introduce an alternative method to estimate $p'_{cell}$.

From the far-field, bubbles can be seen as acoustic point
sources isotropically distributed in space.
In classical scattering theory, pressure fluctuations resulting from such sources are modeled as a stochastic noise.\citep{Ishimaru78,Fouque07}
Following the theory, we model $p'_{cell}$ as white noise that is locally, spatially isotropic.
In that case, $p'_{cell}$ is characterized by its (zero) mean and variance.
In the discretized field, we have the following relation between
$E[p'^2_{cell}]$ and $E[(T_zp'_{cell})^2]$:
\begin{equation}
E[T_zp'^2_{cell}]_{i,j}=
E\left[\frac{1}{n_p}\sum^{n_p}_{k=1}p'^2_{{cell}_{i,j,k}}\right]
\approx
C_T E[p'^2_{cell}]_{i,j},\label{eqn:clt}
\end{equation}
where $C_T=(2N_r+1)/n_p$.
In more general cases where $p'_{cell}$ is modeled as a colored noise, $C_T$ can take a different value.
We note that a method to obtain $C_T$ for a field of spatially isotropic, fluctuating variables was discussed by Brunt et al.\citep{Brunt10}, with applications to isotropic turbulence in the interstellar medium.

Our goal is to recover $p'_{cell}$ at the location of each bubble,
from the value of $T_zp_{cell}$ in a two-dimensional simulation.
To the aim, we estimate $T_zp'^2_{{cell}_{i,j}}$ by sampling the values of
$T_zp_{cell}$ in the neighboring cells:
\begin{equation}
E[(T_zp')^2_{{cell}_{i,j}}]\approx
S[(T_zp'_{{cell}_{i,j}})^2]-S[T_zp'_{{cell}_{i,j}}]^2.
\end{equation}
$S$ is an operator that takes volume-weighted average in a window of cells:
\begin{equation}
S[(\cdot)_{i,j}]=\frac{\sum_{m=i-N_s}^{i+N_s}\sum_{n=j-N_s}^{j+N_s}
(\cdot)_{i,j}V_{{m,n}}}{\sum_{m=i-N_s}^{i+N_s}\sum_{n=j-N_s}^{j+N_s}
V_{{m,n}}},\label{eqn:S}
\end{equation}
where the sampling window is given by $(2N_r+1)\times(2N_r+1)$.
Then we can obtain $E[p'^2_{cell}]_{i,j}$ from $E[(T_zp')^2_{{cell}_{i,j}}]$
using relation (\ref{eqn:clt}).

Rigorous approaches to integrating PDE with stochastic source term (Langevin equation) are available.\citep{Delong13}
Yet, compatibility of such approaches with the other components of the present method, such as the high-order WENO scheme and the stiff dynamics of bubbles, is not guaranteed, and is beyond a focus of the present study.
We therefore solve a deterministic equation by modeling the source as a (smooth) sum of Fourier components with randomized phase\citep{Bechara94}.
Following the method, we express $p'$ as:
\begin{equation}
p'_{i,j}(z,t)=\int A(k)\mathrm{e}^{i(kz-\omega
t+\phi_k)}dk,
\end{equation}
where $k$ is the wave number, $\omega=kc$ is the angular frequency, $A$ is the amplitude, and $\phi_k\in[0,2\pi]$ is the random phase associated with $k$, given a priori.
In the present study, we use a Gaussian spectral power distribution in terms of the wavelength $\lambda=2\pi/k$:
\begin{equation}
A^2(\lambda)=\frac{C_\lambda}{\sigma_\lambda\sqrt{2\pi}}e^{-\frac{(\lambda-\lambda_c)^2}{2\sigma_\lambda^2}},
\end{equation}
where $C_A$ is a normalization constant that satisfies
\begin{equation}
\int A^2(k)dk = E[p'^2_{cell}]_{i,j}.
\end{equation}

Given the physical observation that a dominant structural length scale of $p'$ corresponds to the mean inter-bubble distance, we take $\lambda_c=1/n_b^{1/3}$, where $n_b$ is the local density of the bubble, and $\sigma_\lambda=\lambda_c/2$ as an ansatz.
For numerical representation, we express $p'_{cell}$ as
\begin{equation}
p'_{cell,i,j}(z,t)\approx \sum^{N_\phi}_{i=1}\tilde{A}(k_i)\mathrm{cos}(k_iz-\omega t+\phi_{k_i})\Delta k_i.
\footnote{Note that the energy given by the statistical mean $<p'p'>$ is set equal to the variance: $<p'_{cell}p'_{cell}>_{i,j}=\frac{1}{2\pi}\int_0^{2\pi}p'_{cell,i,j}p'_{cell,i,j}dk=E[p'^2_{cell}]_{i,j}$.}
\end{equation}
$N_\phi=100$ and uniform $\Delta k:\Delta k=k_{max}/N_\phi=\pi/N_\phi\Delta$ give a satisfactory result.
$\phi_{k_i}$ is randomly calculated with $E[\phi_{k_i}]=\pi/2$.

\subsection{Axi-symmetric volume averaged equations}
To model axi-symmetric flows, we define an azimuthal averaging operator $T_\theta$:
\begin{equation}
T_\theta(\cdot)=\frac{1}{2\pi r}\int_0^{2\pi}(\cdot)rd\theta.
\end{equation}
Following the two-dimensional case, we apply $T_\theta$ to
three-dimensional volume averaged equations:
\begin{equation}
\frac{\partial (T_\theta\vector{q}_l)}{\partial t} + \nabla\cdot
\vector{f}(T_\theta\vector{q}_l)
=
\vector{g}(T_\theta\beta,T_\theta\vector{q_l})-\nabla\cdot\vector{f}_{res}
+\vector{g}_{res}\label{eqn:vaeq_symbaxi},
\end{equation}
where
\begin{align}
\vector{g}_{res}
&\approx
T_\theta\left(\beta'\frac{d \beta'}{d t}\right)
+(2T_\theta\beta+1)T_\theta(\beta'\nabla\cdot(\vector{f}-\vector{u}_l\vector{q}_l)').
\end{align}
In order to obtain $\beta'$ and $\dot{\beta'}$, we define a
regularization kernel that maps the volume of bubbles onto the axi-symmetric grid:
\begin{equation}
\delta_{axi}=\frac{1}{2\pi r}\int^{2\pi}_{0}\delta rd\theta
\approx
\frac{1}{2\pi}\sum^{n_p}_{k=1}\Delta\theta_k\delta(d_p(\vector{x}_p,\vector{x}_{i,j,k}),h).\label{eqn:axikernel}
\end{equation}
Here $\vector{x}_{i,j,k}$ is the coordinate of the cell center of finite volume grid cell $I_{i,j,k}$ on three-dimensional cylindrical coordinate:
\begin{equation}
I_{i,j,k}=[z_{i-1/2},z_{i+1/2}]\times[r_{j-1/2},r_{j+1/2}]\times[\theta_{k-1/2},\theta_{k+1/2}],
\end{equation}

where $i$, $j$ and $k$ are the indices in $z-$, $r-$ and
$\theta-$ directions, and $z_{i\pm1/2}$, $r_{j\pm1/2}$ and $\theta_{k\pm1/2}$ are the positions of cell faces.

The total contribution of the volume of bubble $P$ on
$\beta_{{axi}_{i,j}}$ is represented by an overlapping region of an axi-symmetric finite volume cell $I_{i,j}$ with a shape of a cylindrical ring, and the ball within that $V_n$ is smeared over (fig \ref{fig:grids}c).

$p'$ can be recovered in the same procedure as that used for two-dimensional
grids.

For numerical integration of the axi-symmetric volume averaged equations, we spatially discretize equation (\ref{eqn:vaeq_symbaxi}) in the following form:

\begin{equation}
\frac{\partial (T_\theta\vector{q}_l)}{\partial t}
+\frac{\partial \vector{f}^z(T_\theta\vector{q}_l)}{\partial z}
+\frac{\partial \vector{f}^r(T_\theta\vector{q}_l)}{\partial r}
=\vector{s}(T_\theta\vector{q}_l)+\vector{g}(T_\theta\beta,T_\theta\vector{q}_l)+\vector{g}_{res},
\end{equation}
where $\vector{s}$ is the geometrical source term.
This formulation is convenient since we can integrate the equations on
finite volume grid cells defined on two-dimensional Cartesian coordinates\citep{Toro13}:

\begin{equation}
I_{i,j}=[z_{i-1/2},z_{i+1/2}]\times[r_{j-1/2},r_{j+1/2}].
\end{equation}

We note that, special kernel functions similar to (\ref{eqn:2dkernel}) and (\ref{eqn:axikernel}) were previously derived
for smoothed particle hydrodynamics (SPH) to simulate a single phase flow with spherical or cylindrical symmetries
by integrating a standard three-dimensional kernel over the direction of symmetry\citep{Omang06}.

\section{Numerical results}
\label{section:result}

\subsection{Single bubble oscillation}
We verify the method on three-dimensional grids by simulating a single
bubble oscillation under pressure excitation.
First, we consider a bubble with an initial radius of $R_0=50$ $\mu$m excited by
a single cycle of sinusoidal pressure wave with a frequency of $f=150$ kHz and an
amplitude of $p_a=2.0$ atm.
This problem was addressed by FC in order to verify the
model of $p_{out}$ expressed in equation (\ref{eqn:p_out}).
The purpose of the simulation here is to study the effect of WENO
schemes on the radial evolution of the bubble as well as to verify the model of $p_{out}$,
by comparing the results with the analytical solution of the Keller-Miksis equation and the result of FC.
The domain is $x,y,z\in[-10, 10]$ mm.
The flow field is initially ambient and quiescent.
We utilize a $116\times116\times116$ non-uniform computational grid to evolve
the initial condition.
Approximately nonreflecting, characteristic boundary conditions are applied along the domain
boundaries\citep{Thompson87}.
The grid size in the regions around the bubble: $x\in[-5,
5]$ mm, is uniform with $\Delta_x = \Delta_y = \Delta_z=100$
$\mu$m.
The bubble is located at the origin. A planer acoustic source is placed at
$x=-1$ [mm] to send the pressure wave in $+x$ direction.
\begin{figure}
 \begin{subfigmatrix}{1}
  \subfigure[]{\includegraphics[width=81mm]{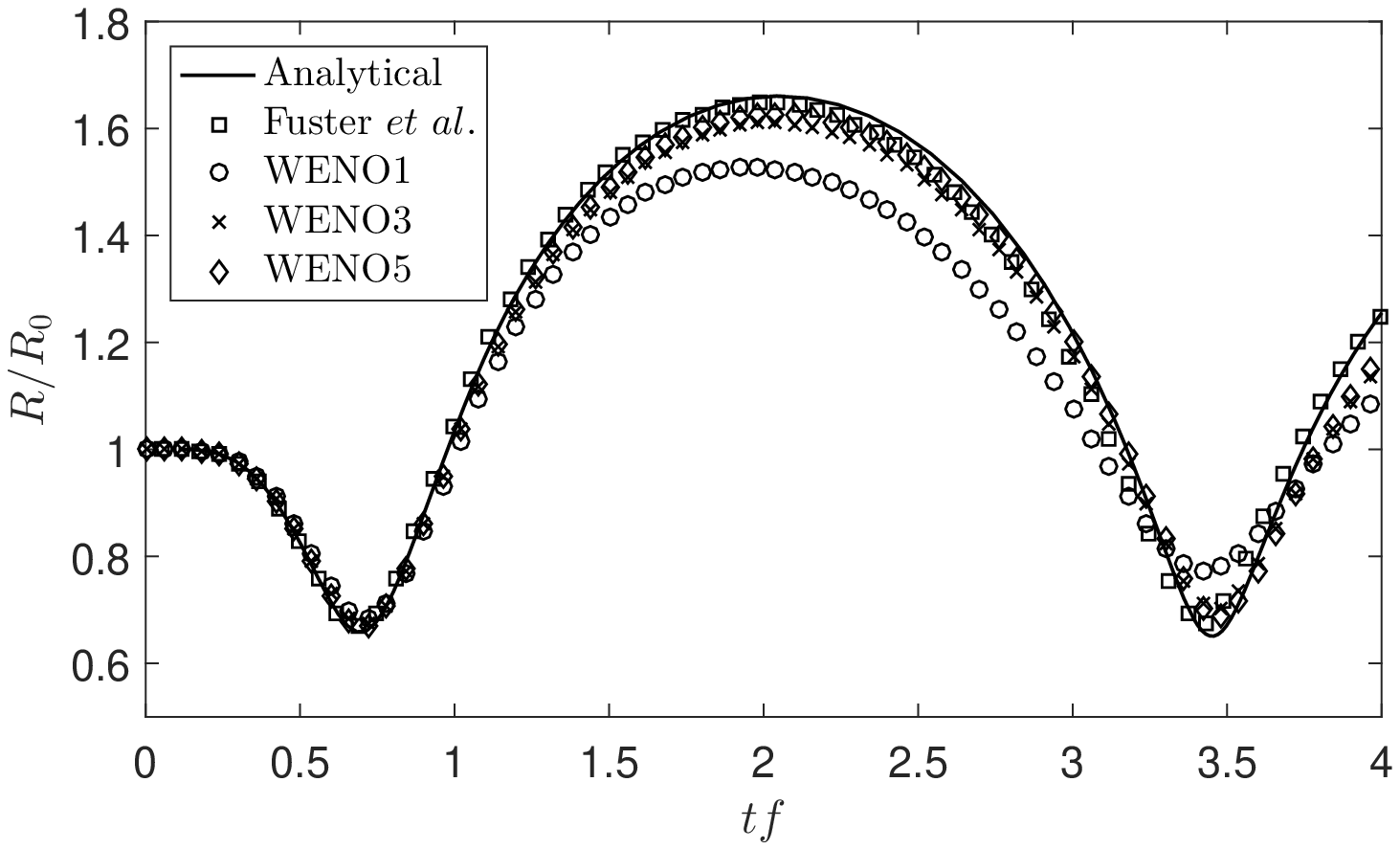}}
  \subfigure[]{\includegraphics[width=81mm]{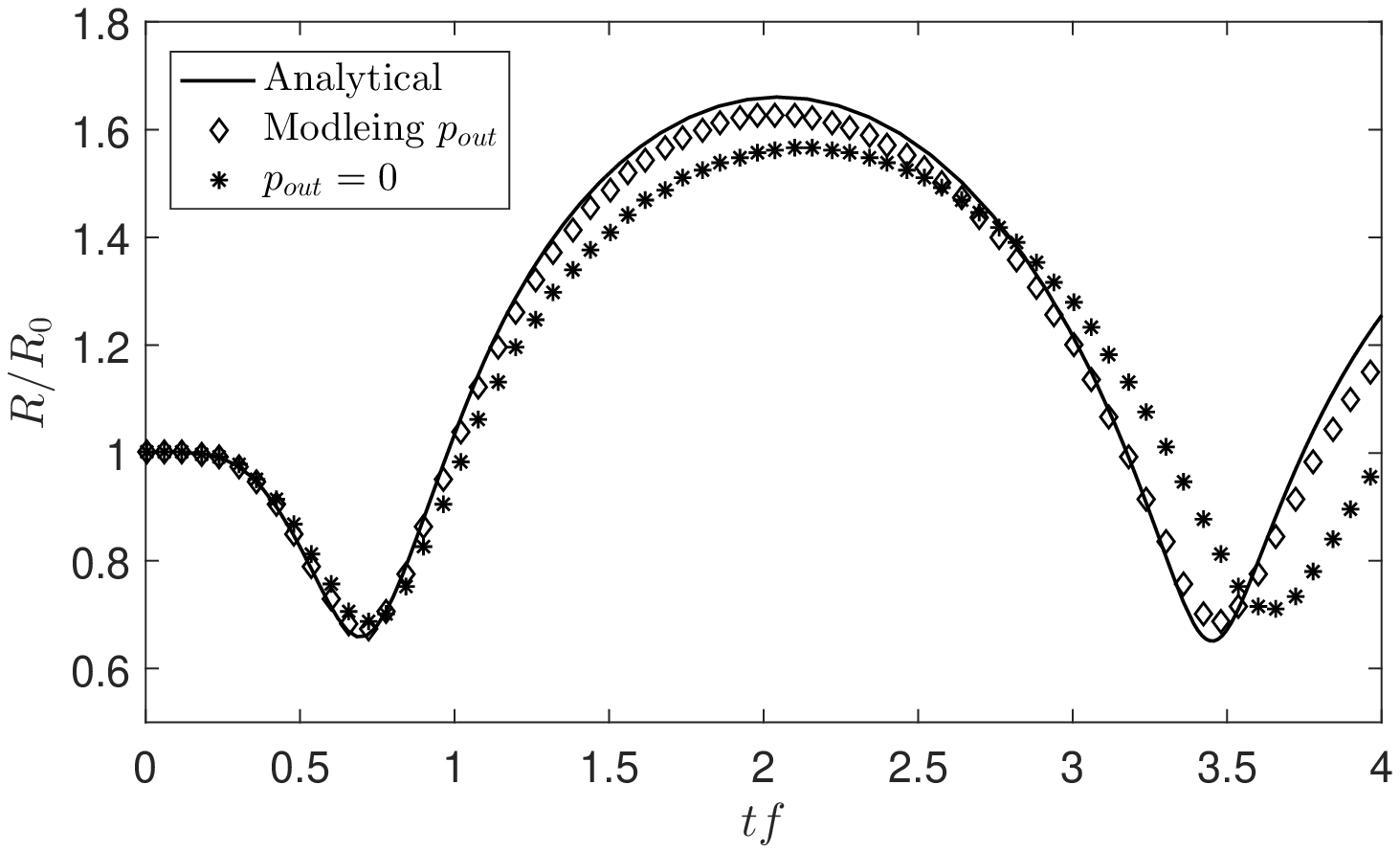}}
  \end{subfigmatrix}
  \caption{Evolution of a single, isolated bubble with an initial radius
  of 50 $\mu$m under excitation by a single cycle of a sinusoidal pressure wave with a
  frequency of $150$ kHz and an amplitude of 2 atm, as a function of the non-dimensional time $tf$.
  (a) Results of FC and the present study using various orders of WENO are compared. (b) Results using WENO5 with and without modeling $p_{out}$ are compared.}
  \label{fig:rpa}
\end{figure}
In Fig. \ref{fig:rpa}a we are comparing results of the present method with
various orders of WENO scheme, the analytical solution, and FC.
The amplitude of oscillation predicted by WENO1 is slightly lower
than those of analytical solution and FC, while the higher-order
methods give satisfactory results.
The discrepancy from the result of FC and the improvement by using a higher order WENO can be explained by the dissipative property of WENO.
FC employed a non-dissipative numerical method\citep{Honein04}.
Compared to such solvers, WENO-based schemes are inherently
dissipative\citep{Pirozzoli06}, but stable for capturing shocks and materials interfaces \citep{Coralic14}.
In Fig. \ref{fig:rpa}b we compare results of WENO5 with and without the model
of $p_{out}$.
With the model, the numerical solution agrees with the analytical solution,
while without it, both the amplitude of the oscillation and the timing of the
second rebound deviate from the analytical solution.

Next, we simulate the dynamics of a bubble with an initial radius of
$R_0=10$ $\mu$m excited by a single cycle of a sinusoidal pressure wave with a frequency of $f=300$ kHz and an
amplitude of $p_a=1$ atm.
We use the same simulation domain, boundary conditions and acoustic source as
the previous case. We track the evolution of the pressure at $[-1.0, -1.0,
-1.0]$ mm during simulations that are evolvefd with various grid spacings in order to assess the effect of the grid size on the pressure waves scattered by
the bubble.
\begin{figure}
 \begin{subfigmatrix}{2}
  \subfigure[]{\includegraphics[width=97mm]{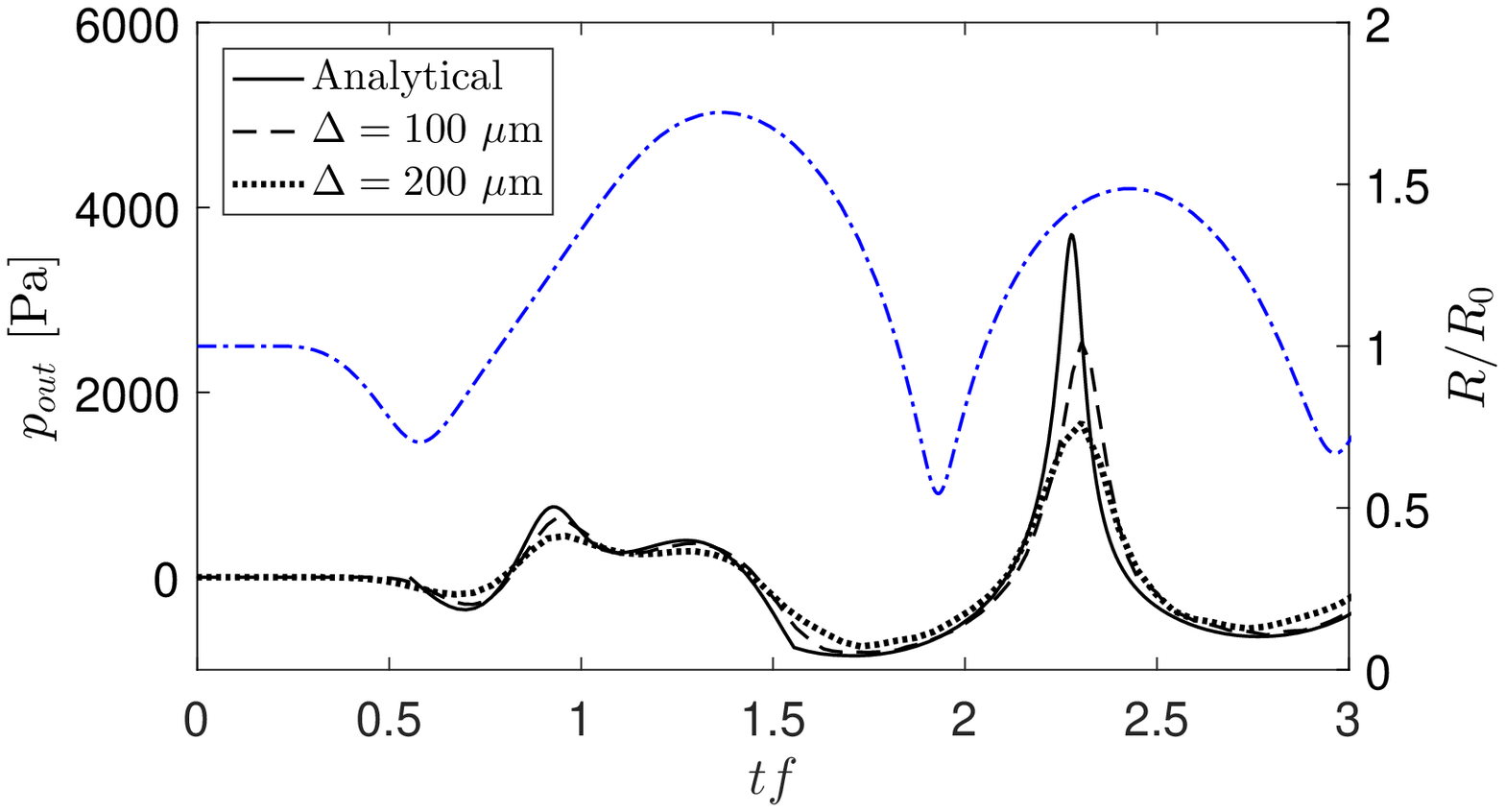}}
  \subfigure[]{\includegraphics[width=62mm]{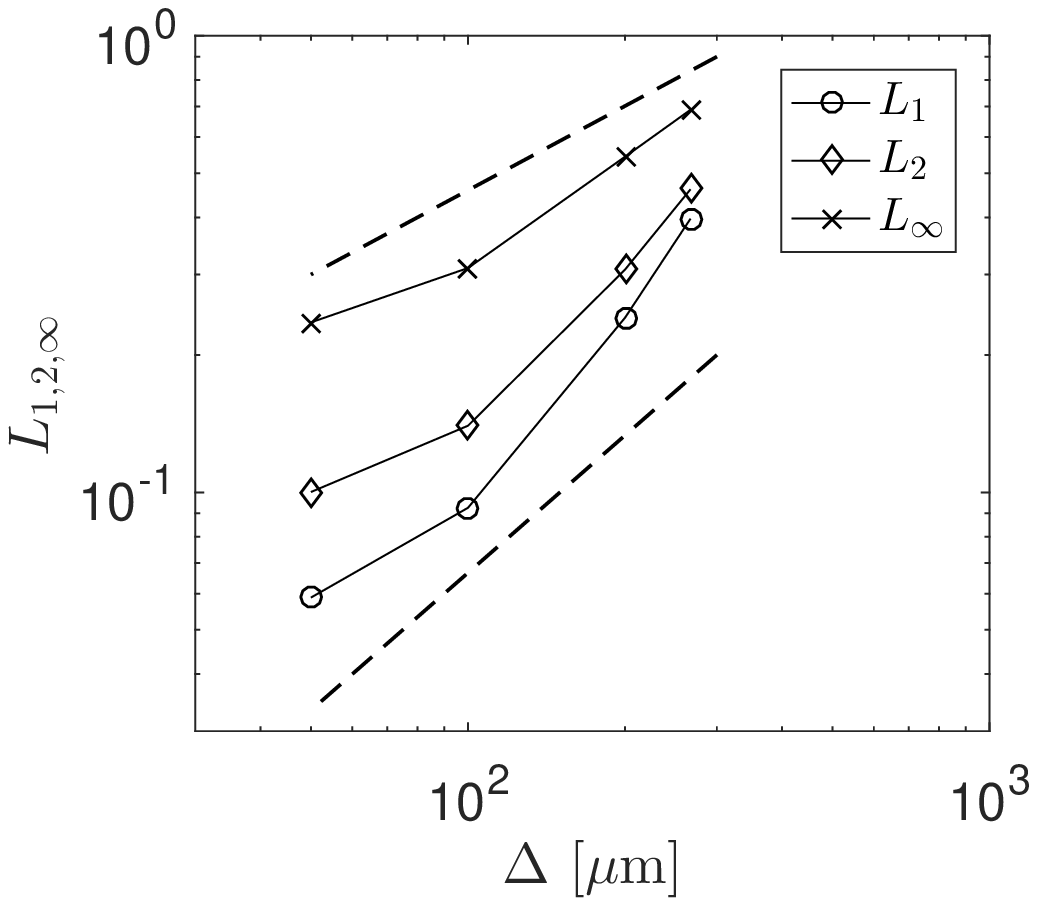}}
  \end{subfigmatrix}
  \caption{Scattered pressure wave from a single, isolated bubble with an
  initial radius of 10 $\mu$m under excitation with a single cycle of sinusoidal pressure wave
  with a frequency of $300$ kHz and amplitude of 1 atm.
  (a) Evolution of the pressure at [-1.0, -1.0, -1.0] mm.
  Results using a grid size of $\Delta=100$ and $200$ $\mu$m in the bubble are
  compared with analytical solution.
  Analytical solution of the evolution the bubble is also plotted. (b) The error norm as a function of the grid size. Reference slopes for first- and half-order convergence are included.
  }
  \label{fig:rp_pres}
\end{figure}
Fig \ref{fig:rp_pres}a shows the results using $\Delta=100$ and $200$ $\mu$m
in the region of the bubble, and an analytical solution derived by solving
the Keller-Miksis equation. The pressure evolution is captured well, even with
the grid size much larger than the bubble size.
The simulated value of the peak pressure, due to the second collapse of the
bubble, approaches the analytical value on the finer grid.
Fig \ref{fig:rp_pres}b shows the error
\begin{equation}
L_{n}=\frac{[\int_0^3 |p(t)|^n dt]^{1/n}-[\int_0^3 |p_{KM}(t)|^n
dt]^{1/n}}{[\int_0^3 |p_{KM}(t)|^n dt]^{1/n}}
\end{equation}
for $n=1$,2 and $\infty$, where $p_{KM}$ denotes the analytical solution derived using the Keller-Miksis equation.
The results demonstrate convergence, but we note that some saturation of the error is evident.
This may be related to errors in the Keller-Miksis solution which makes a weakly-compressible assumption for the liquid.
Nevertheless, the results shown in this section confirm that radial evolutions of
a bubble as well as the bubble-scattered pressure waves are correctly
captured.

\subsection{Bubble screen problem}
In order to verify the reduced model that uses two-dimensional volume-averaged equations, we simulate interactions of a bubble screen with a single cycle of plane, sinusoidal pressure wave using both the two- and three-dimensional models, and compare the results.
\begin{figure}
  \center
  \includegraphics[width=160mm,trim = 00mm 0mm
  00mm 0mm,clip]{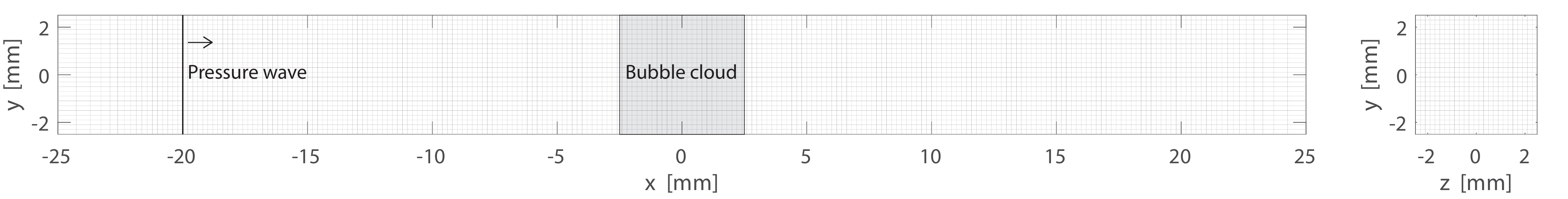}
  \caption{Schematic of the initial condition and the three-dimensional
  computational grid (only one of every two cells shown) for the bubble screen problem.}
   \label{fig:screen} 
\end{figure}

Fig. \ref{fig:screen} shows the schematic of the simulation setup.
The domain is $x\in[-250, 250]$, $y,z\in[2.5,-2.5]$ mm.
We utilize a $572\times50\times50$ and $572\times50$ non-uniform computational grids for three-dimensional and two-dimensional simulation to evolve the initial condition, respectively.
A periodic boundary condition is applied along the domain boundaries
perpendicular to the $y$ and $z$ axes.
Non-reflective boundary conditions are implemented on the boundaries
perpendicular to the $x$ axis. Grid is smoothly stretched away from the bubble screen to prevent contamination by reflections. The grid in the bubble screen region, $x\in[-25,
25]$ mm, is uniform with $\Delta_x = \Delta_y = \Delta_z=100$
$\mu$m.
The region of bubble screen is $x,y,z\in[-2.5,2.5]$ [mm].
Bubbles with an initial radius of 10 $\mu$m are randomly, homogeneously distributed in the region of the screen,
with a given initial void fraction, $\beta_0$.
The flow field is quiescent and at ambient pressure at the initial condition.
A planer acoustic source is located at $x=-25$ [mm] to excite a single cycle of
a sinusoidal pressure wave with an amplitude of 1 MPa and a frequency of 300 kHz
in $+x$ direction.
The resulting bubble oscillations are nonlinear and distinct from the results of classical bubble screen problems that considers excitations of linear oscillations of bubbles using a weak pressure wave\citep{Commander89}.

\begin{figure}
\center
 \begin{subfigmatrix}{1}
  \subfigure[]{\includegraphics[width=80mm]{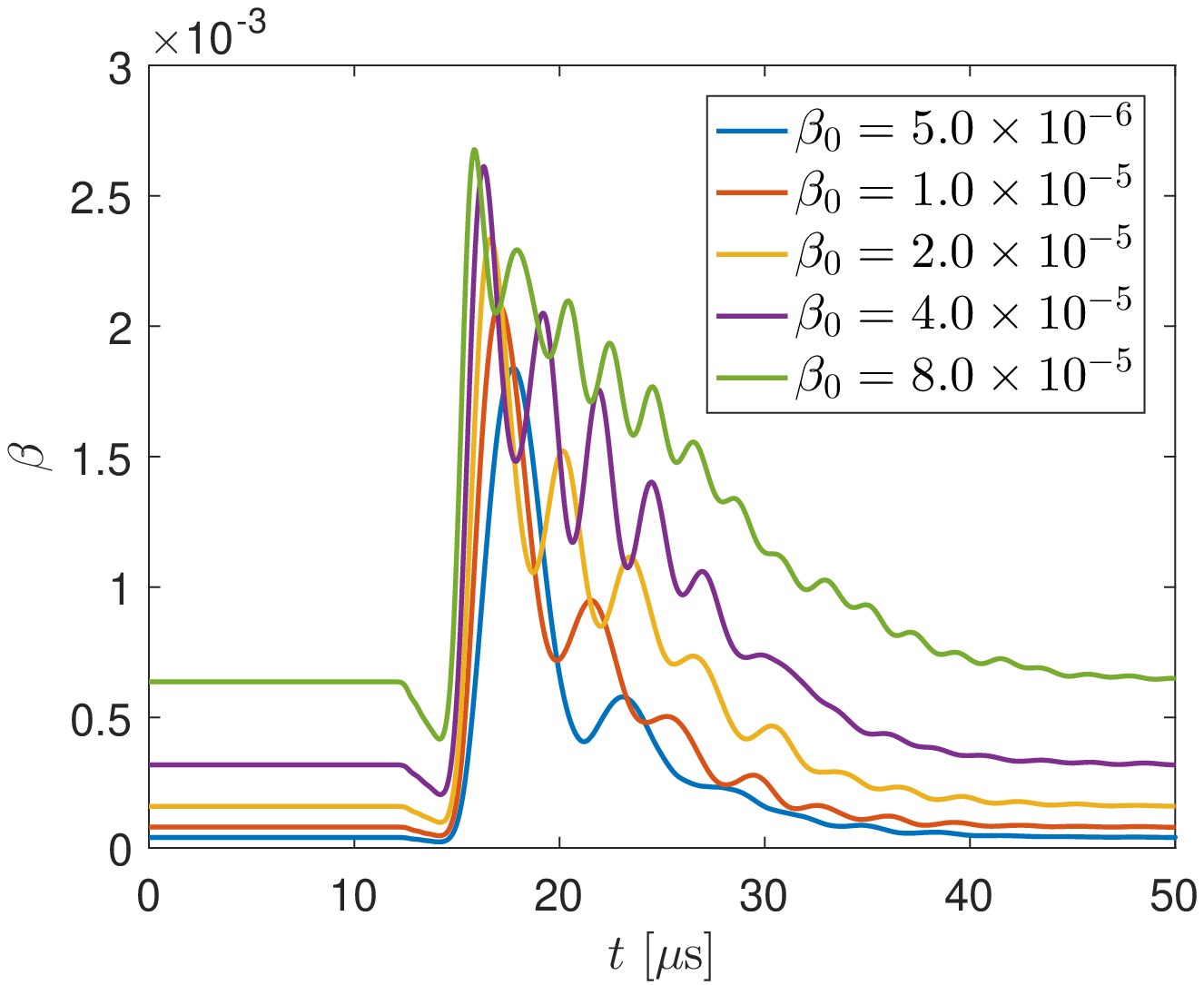}}
  \subfigure[]{\includegraphics[width=80mm]{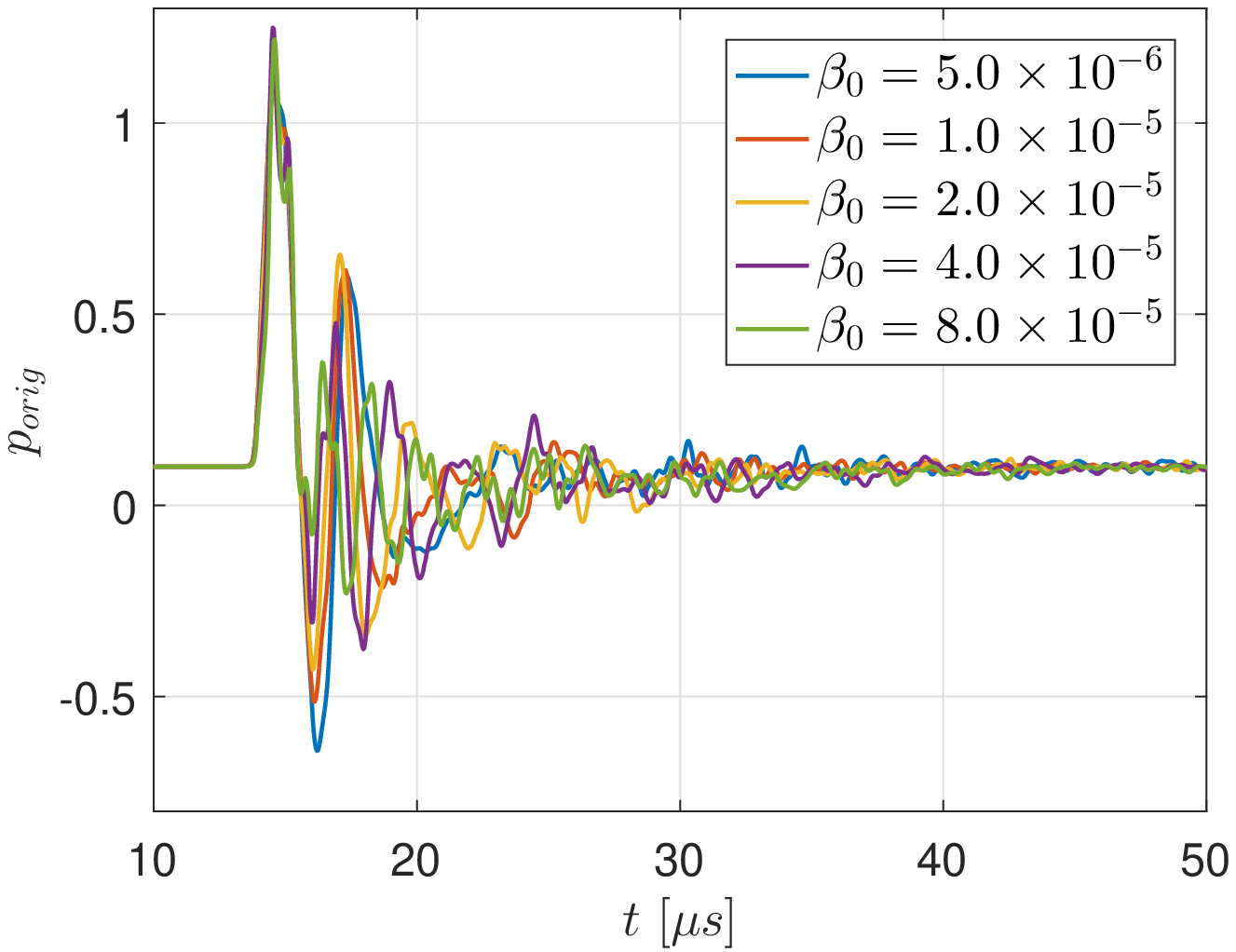}}
  \end{subfigmatrix}
  \caption{Evolution of (a) averaged void fraction and (b) maximum bubble
  radius. during an interaction of a bubble screen and a single cycle of plane,
  sinusoidal pressure wave. Results on three-dimensional grids,
  two-dimensional grids with/without the stochastic closure are compared.}
   \label{fig:void5} 
\end{figure}

Fig \ref{fig:void5}a shows the time evolution of the void fraction of the screen with various values of $\beta_0:\beta_0=[0.5,1.0,2.0,4.0,8.0]\times 10^{-5}$.
For all cases, $\beta$ rapidly grows after the passage of the wave, then smoothly decays with oscillations. The oscillations are induced by reverberations of the pressure waves trapped inside the screen. In fig \ref{fig:void5}b, we plot the time evolution of the pressure at the origin during the same set of simulations. $p_{orig}$ grows and decays during the passage of the wave, then presents rapid fluctuations induced by the oscillations of surrounding bubbles.

\begin{figure}
  \center
  \includegraphics[width=120mm,trim = 00mm 0mm
  00mm 0mm,clip]{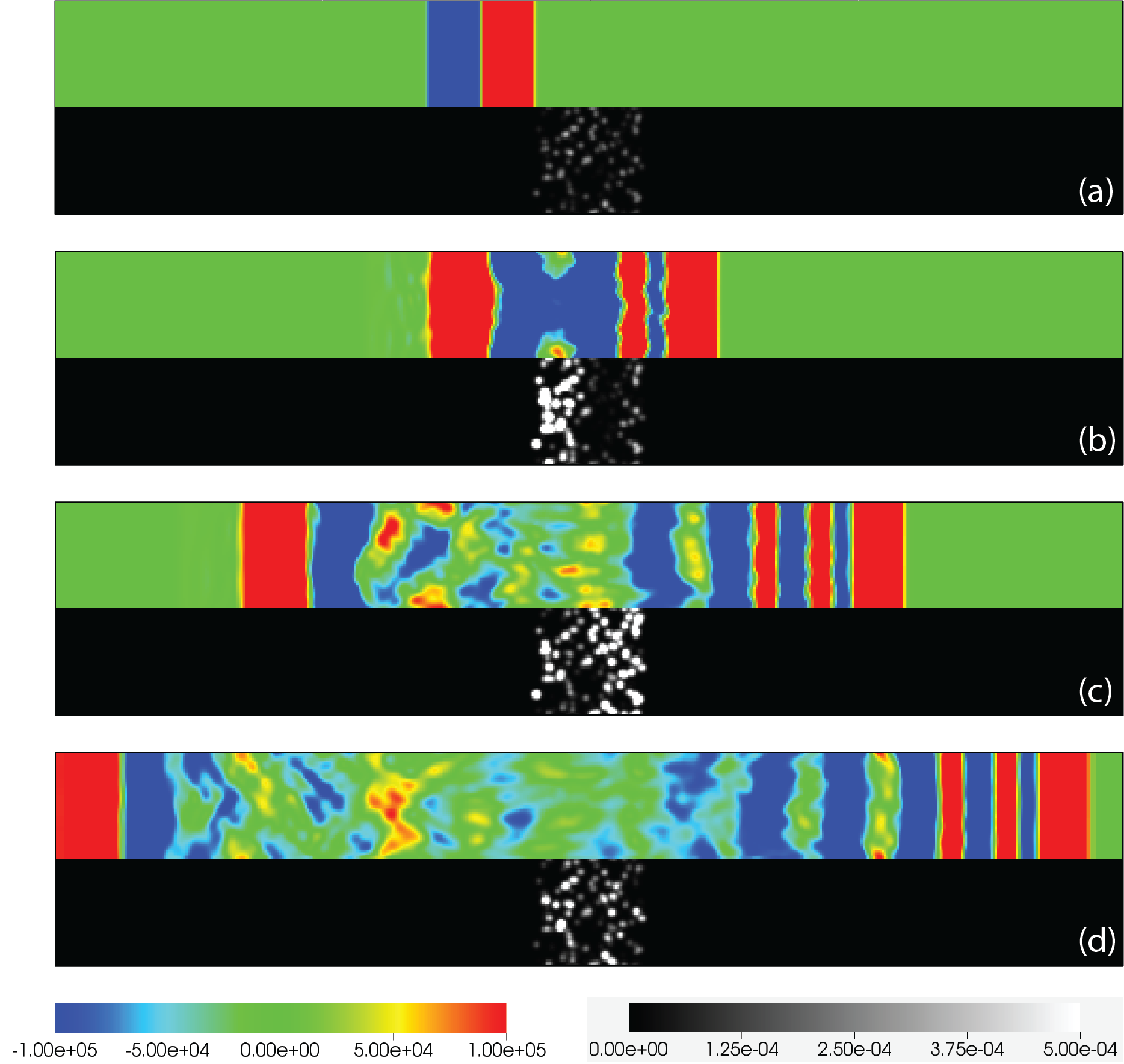}
  \caption{Snapshots of the pressure (top half) and the void fraction (bottom half) contours on $x-y$ plane in the range of $x\in[-25, 25]$ and $y\in[-2.5, 2.5]$ mm for three-dimensional simulation with $\beta_0=4.0\times 10^{-5}$. Note that the pressure is plotted at levels much smaller than the amplitude of the initial wave in order to highlighted the bubble-generated and scattered fields. (a) $t=12$ $\mu$s, (b) 18 $\mu$s, (c) 24 $\mu$s and (d) 30 $\mu$s, respectively.}
   \label{fig:snapshots} 
\end{figure}

Fig \ref{fig:snapshots} shows the time evolution of the pressure and the void fraction contours on $x-y$ plane for the case with $\beta_0=4.0\times 10^{-5}$.
The pressure wave is partially reflected by the screen.
The tensile part of the wave causes growth of the bubbles and subsequent radial oscillation.
The complex structure of the scattered waves is due to the oscillation of bubbles that last longer than the passage of the wave.

In order to verify relation (\ref{eqn:clt}), we compute $E[p'^2_{cell}]/E[(T_zp'_{cell})^2]$ on $x-y$ plane by post-processing the three-dimensional simulation data at each cell and take an average of the values over the region of $x,y\in[-2.0,2.0]$ mm to obtain the empirical values of $C_T$:
\begin{equation}
C_{T_{Emp}}= \frac{1}{L_xL_y}\int_{-2}^2\int_{-2}^2 \frac{E[p'^2_{cell}]}
{E[(T_z p'_{cell})^2]} dxdy,
\end{equation}
where $L_x=L_y=4$ mm.

\begin{figure}
\center
 \begin{subfigmatrix}{1}
  \subfigure[]{\includegraphics[width=80mm]{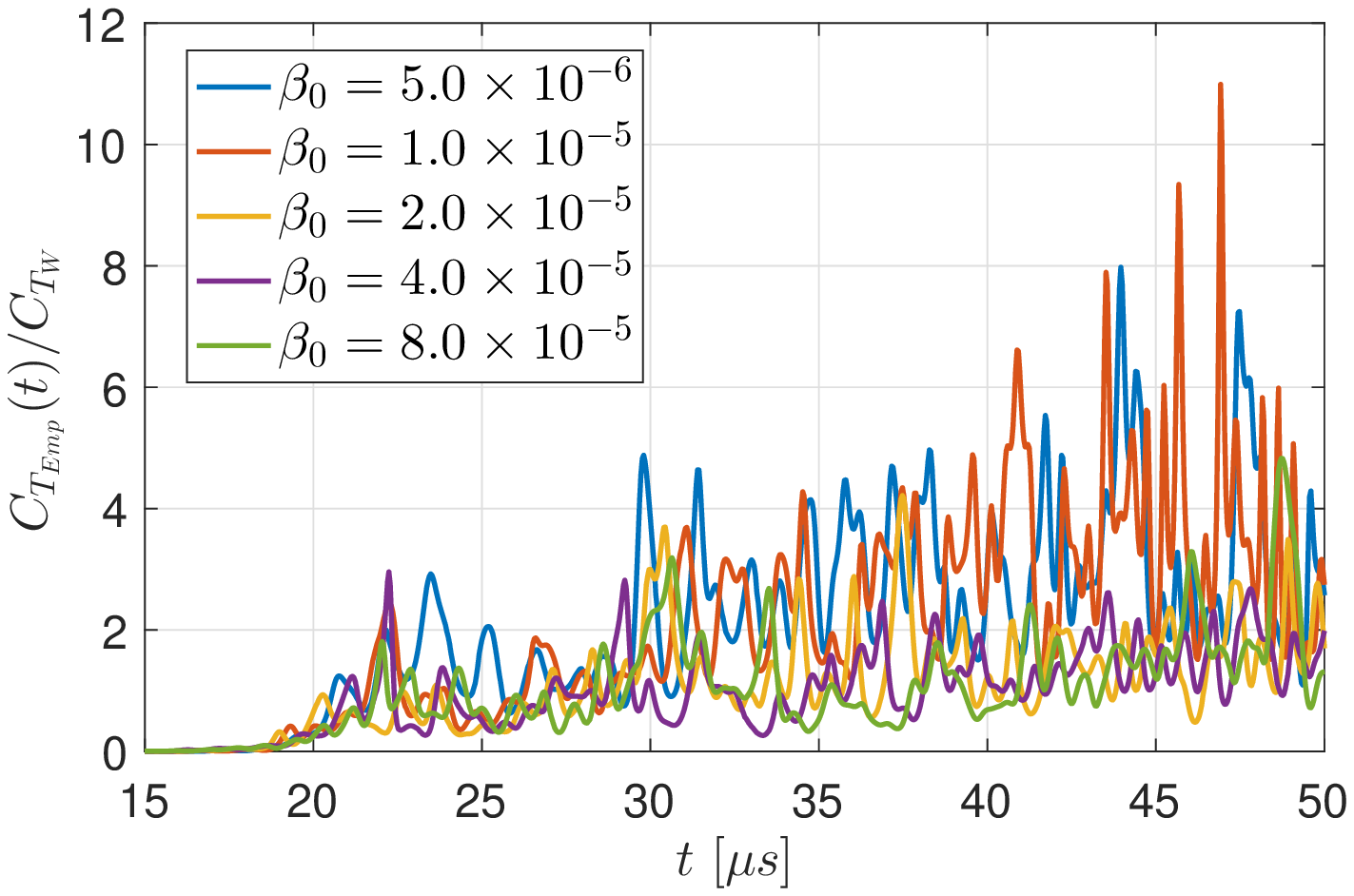}}
  \subfigure[]{\includegraphics[width=80mm]{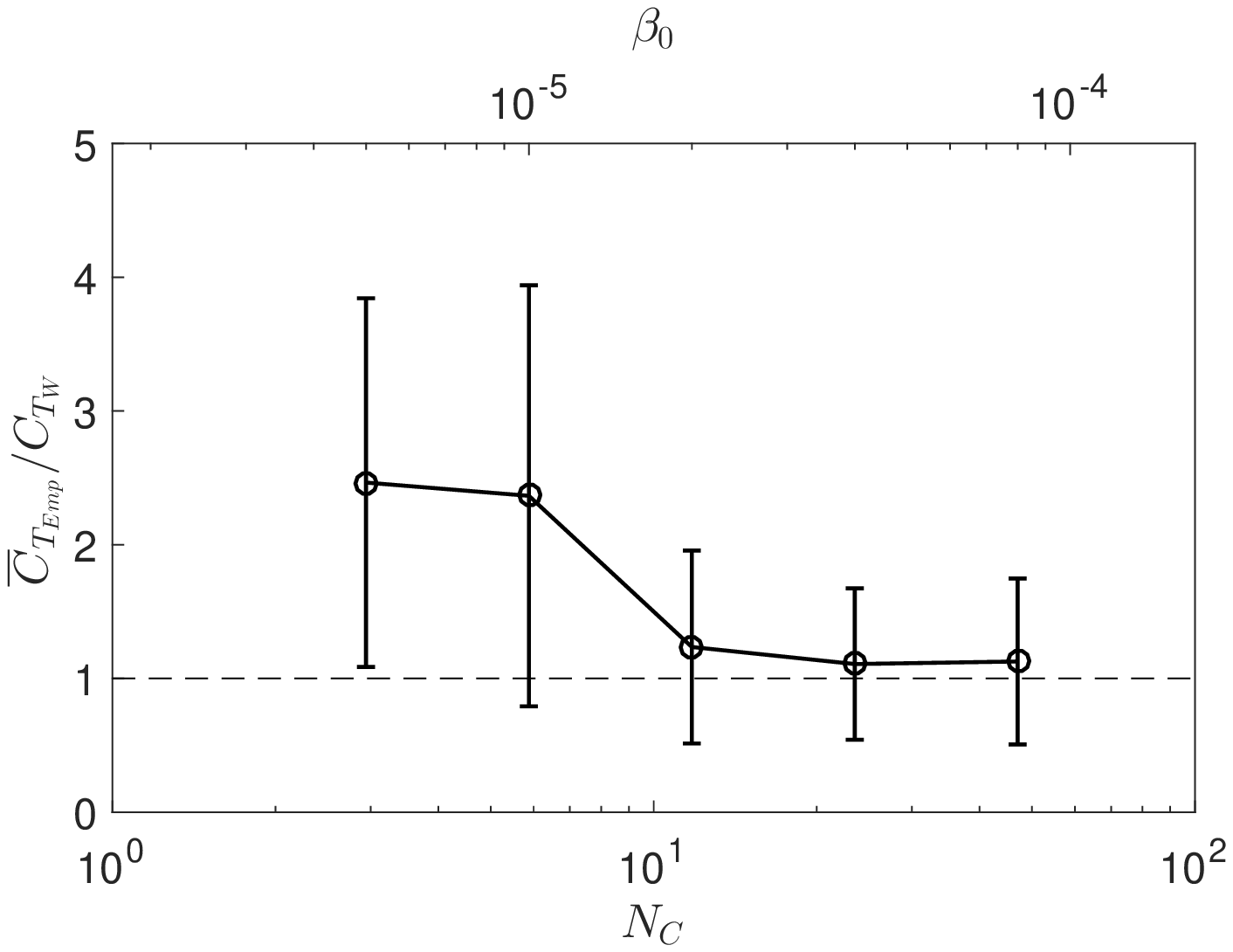}}
  \end{subfigmatrix}
  \caption{(a)Evolution of $C_{T_{Emp}}/C_{T_W}$ obtained from the three-dimensional simulations with $\beta_0=[0.5,1.0,2.0,4.0,8.0]\times10^{-5}$.              (b)$\overline{C}_{T_{Emp}}/C_{T_W}$ obtained from the same simulations, as a function of $\beta_0$ and $N_C$.}
   \label{fig:noise} 
\end{figure}

Fig \ref{fig:noise}a shows the time evolution of the ratio of $C_{T_{Emp}}$ to its theoretical value obtained from the white noise model $C_{T_w}=C_T$, for simulations with various $\beta_0$.
For all cases of $\beta_0$, $C_{T_{Emp}}=0$ until around at $t=20 \mu$s, since there are no pressure fluctuations along the $z$ axis in the domain.
After the passage of the wave, $C_{T_{Emp}}$ grows due to the bubble dynamics.
In the cases with $\beta_0=5.0\times 10^{-6}$ and $1.0\times 10^{-5}$, $C_{T_{Emp}}/C_{T_w}$ fluctuates rapidly and grows to a value of 10, while in the cases with higher values of $\beta_0$, the fluctuation is smaller and the value stays close to 1. Fig \ref{fig:noise}b shows the averaged value of $C_{T_{Emp}}/C_{T_w}$ within the interval of $t=[20,50]$ $\mu$s as a function of $\beta_0$ and $N_C$, where $N_C$ is the averaged number of bubbles contained in the region that the operator $S$ averages over (see equation (\ref{eqn:S})).
In accordance with fig \ref{fig:noise}a, with $\beta_0=5.0\times 10^{-6}$ and $1.0\times 10^{-5}$, $\overline{C}_{T_{Emp}}/C_{T_w}$ takes a value much larger than 1, while with $\beta_0$ higher than $2.0\times 10^{-5}$ it takes a value close to 1.
This transition corresponds to the value of $N_C$ exceeding $O(10)$.
The results indicate that with $N_c<O(10)$ the distribution of $p'_{cell}$ is not locally isotropic in the averaging window, while with $N_c>O(10)$, the distribution of $p'_{cell}$ becomes locally isotropic and $p'_{cell}$ is well modeled by white noise, and thus relation (\ref{eqn:clt}) holds.

In order to assess the improvement by the stochastic closure used for the two-dimensional volume-averaged equations,
we simulate the bubble screen problem with $N_{ens}=15$ distinct initial distributions of bubbles in the screen, with a fixed value of $\beta_0:\beta_0=4.0\times10^{-5}$, using the three-distinct models: the three-dimensional model, two-dimensional model, and two-dimensional model with $p'_{cell}=0$, respectively.
Then, for each method, we empirically obtain the ensemble-averaged solution by averaging the results of simulations:
\begin{equation}
<f(\vector{x},t)>=\frac{1}{N_{ens}}\sum_{i=1}^{N_{ens}} f_i(\vector{x},t),
\end{equation}
where $f_i$ is an arbitrary quantity computed in $i$-th simulation and $<\cdot>$ denotes the ensemble average.
The purpose of comparing the ensemble-averaged solutions is to eliminate the incoherence among the distinct simulations that originates from the differences in the spatial distributions of bubbles, so that we can isolate the effect of the differences in the models on the resulting solutions.

\begin{figure}
\center
 \begin{subfigmatrix}{1}
  \subfigure[]{\includegraphics[width=80mm]{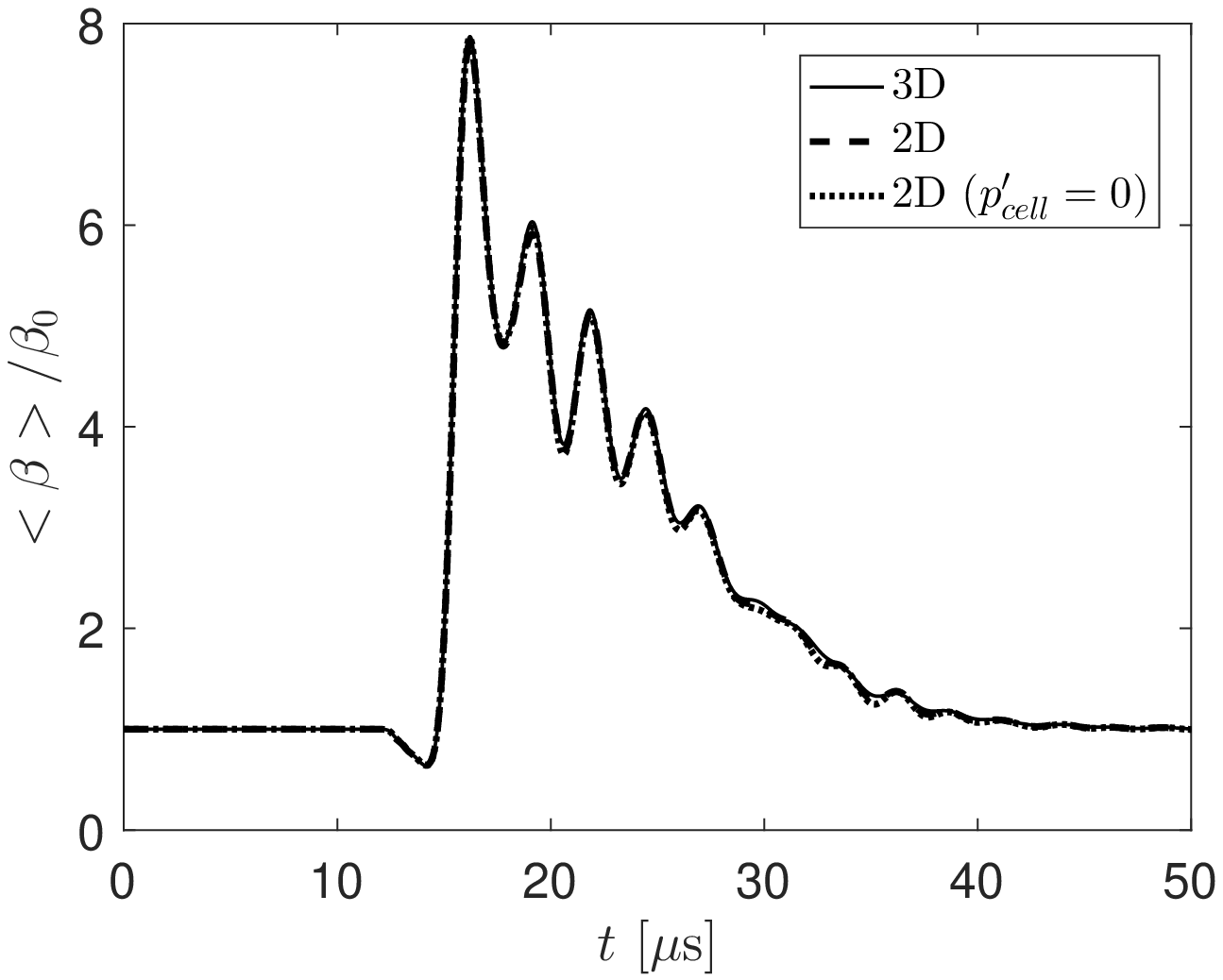}}
  \subfigure[]{\includegraphics[width=80mm]{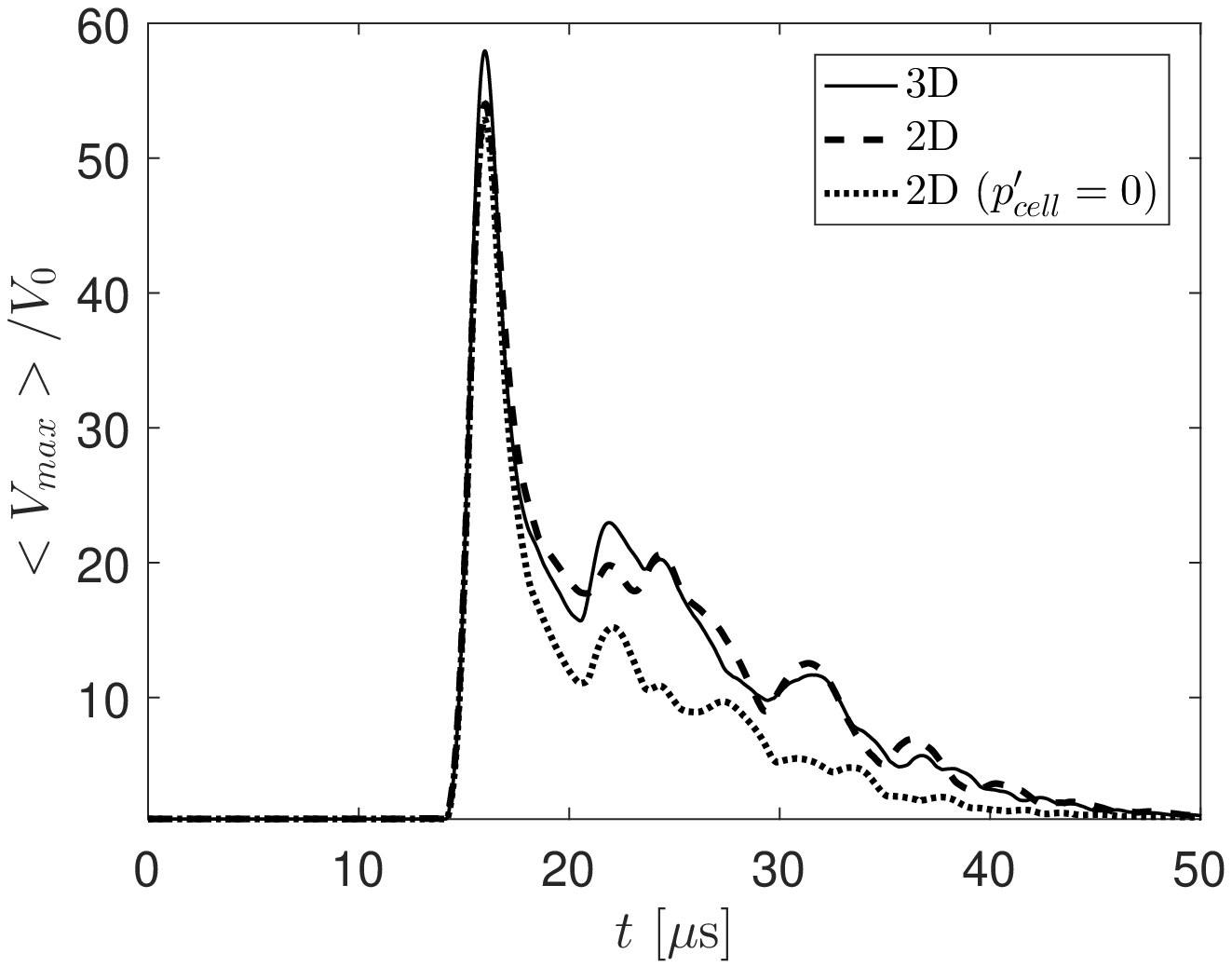}}
  \end{subfigmatrix}
  \caption{Evolution of the ensemble averaged values of (a)void fraction and (b) maximum bubble radius in the screen. Results using three-dimensional model, two-dimensional model and two-dimensional model with $p'_{cell}=0$ are compared.}
   \label{fig:3d2d} 
\end{figure}

In fig. \ref{fig:3d2d} we plot the evolution of the ensemble-averaged values of the void fraction and those of the normalized maximum radius of bubbles during the simulations.
Interestingly, the choice of the model makes no visible difference in the void fraction.
Meanwhile, the maximum bubble radius is significantly under-estimated after $t=20$ $\mu$s unless the closure is applied.

\begin{figure}
\center
  \includegraphics[width=80mm]{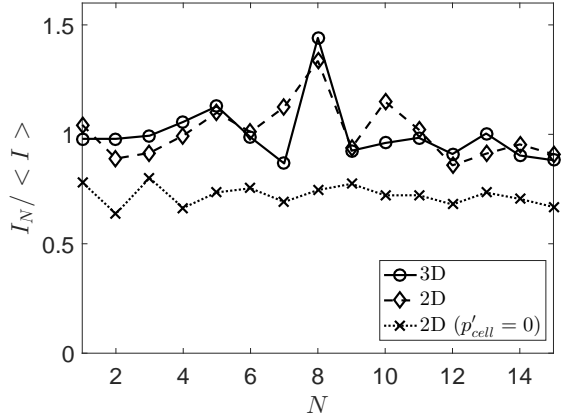}
  \caption{Fluctuations in $I_N$ through 15 simulations with distinct initial conditions of bubbles: $I_N/<I>:N\in[1,15]$.
  Results using three-dimensional model, two-dimensional model and two-dimensional model with $p'_{cell}=0$ are compared.}
   \label{fig:sym2} 
\end{figure}

Fig. \ref{fig:sym2} shows the time integral of $V_{max}$ within the interval of $t=[0,50]$ $\mu$s obtained from the each solution of the $N_{ens}$ simulations, normalized by its ensemble averaged value $<V_{max}>$ obtained from the three-dimensional model:
\begin{equation}
I_N=\int_0^{t_f} V_{N,max} dt,
\end{equation}
where subscript $N$ denotes $N$-th simulation.
Both the value of $I_N$ and the magnitude of fluctuations among distinct simulations are smaller in the two-dimensional simulations with $p'_{cell}=0$, compared to those in the three-dimensional simulations. Meanwhile, the result obtained with the two-dimensional model with $p'_{cell}$ agrees the three-dimensional simulations relatively well.
We also note that with all the models $<I>$ converges with $N_{ens}=15$, and the ensemble averaged values obtained with the two-dimensional model with the closure is sufficiently close to that of the three-dimensional model, while that obtained without closure is smaller by 28\%.
Particularly in the present test case, the sub-grid pressure fluctuations influence only the maximum volume, but not the averaged void fraction.
A physical interpretation of this result is that, even though the mean response of the volume of bubbles to the sub-grid pressure fluctuations is statistically close to zero, the coherence in the volumetric oscillations of the bubbles is lowered by the action of the random noise, and therefore the local maximum of the volume of bubbles is increased.

\subsection{Cloud cavitation in a high-intensity ultrasound wave}
Lastly, we simulate interactions of a spherical bubble cloud with planer, multiple-cycles of a sinusoidal pressure wave using the three-dimensional and axi-symmetric models and compare the results with a high-speed image of a cavitation bubble cloud obtained in an experiment.
The purpose of this case is to further demonstrate the feasibility of the proposed method for cloud cavitation in ultrasound-based lithotripsy, where we observe a bubble cloud with a size of $O(1)$ mm interacting with ultrasound waves with a frequency of $O(0.1-1)$ MHz and amplitude of $O(1-10)$ MPa\citep{Maeda15}.
This problem is particularly challenging for previous approaches since the wavelength of the incident pressure wave is close to the size of the bubble cloud, thus the pressure field needs to be resolved at a scale smaller the cloud. Moreover, the amplitudes of the bubble-scattered pressure waves are strong so that a fully compressible liquid is needed.
\begin{figure}
\center
 \includegraphics[width=70mm]{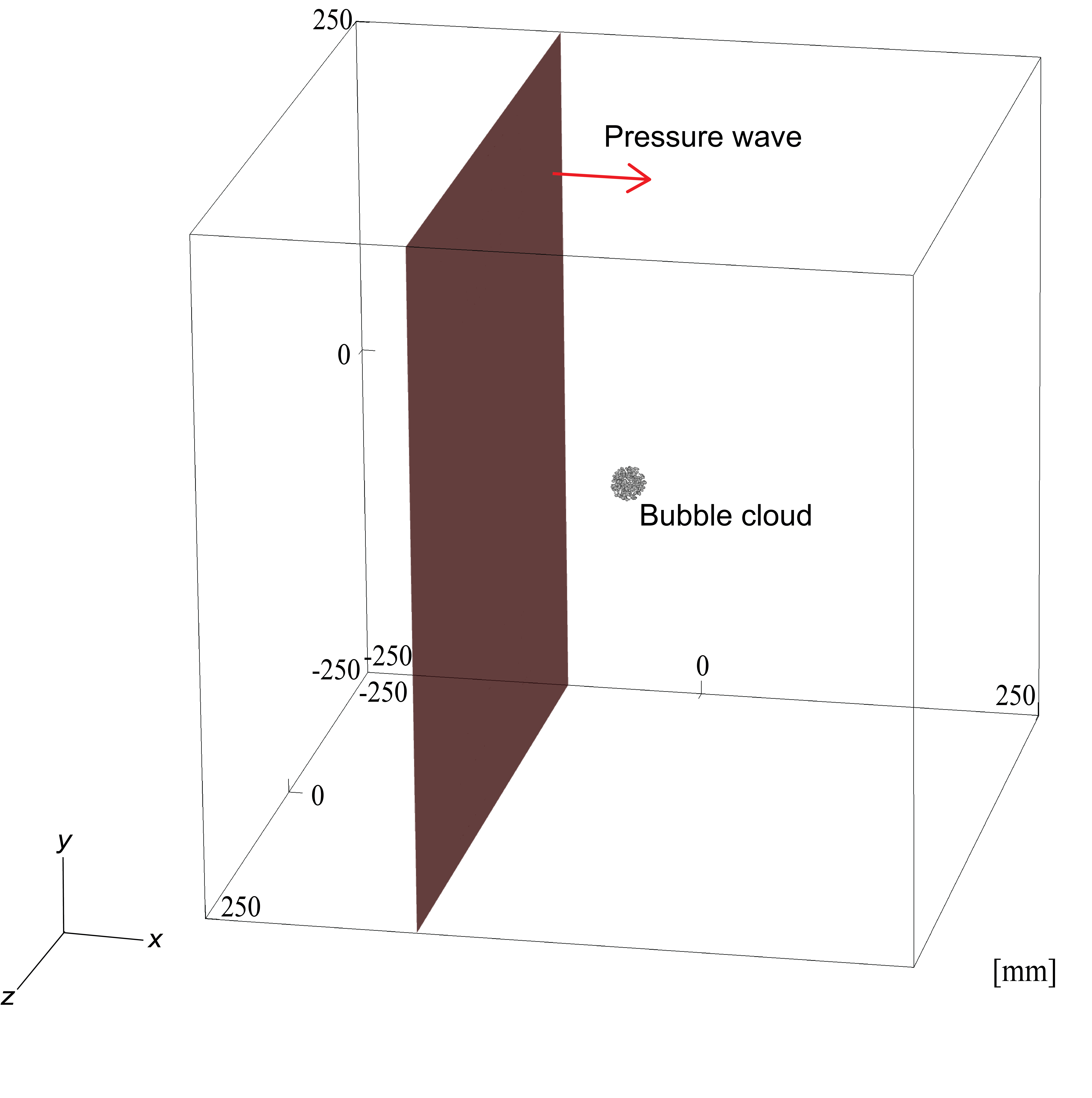}
  \caption{Schematic of the simulation setup for the wave-cloud interaction
  problem. We excite a planer, 10 cycles of a sinusoidal pressure wave with an
  amplitude of 1.0 MPa and a frequency of 300 kHz from a source plane located at
  $x=-20$ mm in the $+x$ direction. A bubble cloud with a radius of 2.5 mm is
  located at the origin. 625 Bubbles are randomly distributed in the region of
  the cloud. The radii of bubbles follow Gaussian distribution with a mean of 10 $\mu$m and a standard deviation of $2.5$ $\mu$m. The initial void fraction of the cloud is $\beta_0=4.87\times10^{-5}$.}
  \label{fig:cloud3d} 
\end{figure}
Fig. \ref{fig:cloud3d} shows the schematic of the simulation setup.
In the three-dimensional simulation, the simulation domain is $x,y,z\in[-250, 250]$ mm.
The bubble cloud with a radius of 2.5 mm is located at the origin, immersed in water.
The pressure of the domain is uniformly ambient and the flow field is quiescent at the initial condition.
A planer acoustic source is located at $x=-20$ [mm] to send 10 cycles
of sinusoidal pressure waves with a frequency of 300 kHz and an amplitude of 1
MPa in $+x$ direction toward the cloud.
The pressure wave begins with compression and ends with tension.
We utilize a $572\times572\times572$ non-uniform computational grid to evolve
the initial condition.
Non-reflective boundary conditions are applied along the domain boundaries.
The grid size in the regions around the bubble: $x,y,z\in[-25,
25]$ mm, is uniform with $\Delta_x = \Delta_y = \Delta_z=100$
$\mu$m.
Grid is smoothly stretched away from the wave-cloud interaction region to prevent pollution from reflections.
625 bubbles are randomly distributed in the cloud.
The radii of bubbles are selected from a Gaussian distribution with a mean of 10 $\mu$m and a
standard deviation of $2.5$ $\mu$m. The initial void fraction of the cloud is $4.87\times10^{-5}$.
In the axi-symmetric simulation, the simulation domain is $z\in[-250, 250]$ and $r\in[0, 250]$ mm.
Grid size and stretching on $r-z$ plane follow those on $x-y$ plane in the three-dimensional simulation.
The same initial condition of bubbles is used as the three-dimensional simulation.
Various other parameters in the simulations are chosen to match experiments.
A discussion of the details is beyond the scope of this paper and omitted here.

\begin{figure}
\center
 \includegraphics[width=125mm]{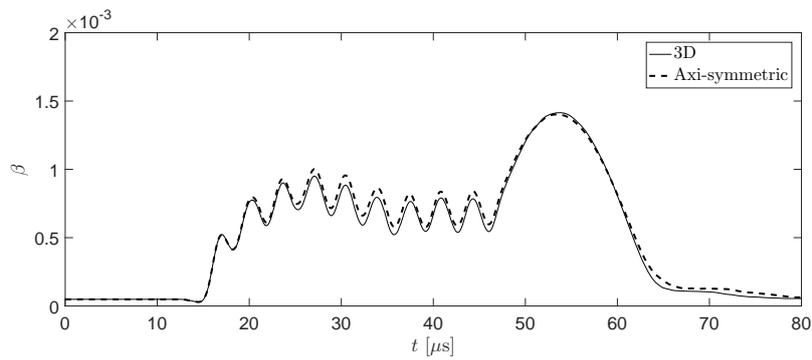}
  \caption{Evolution of the void fraction during cloud-wave interaction.}
  \label{fig:cloud3d2d} 
\end{figure}

Fig. \ref{fig:cloud3d2d} shows the time evolution of the void fraction.
The results obtained with the two models agree very well with each other.
The pressure front reaches the surface of the cloud at $t=15.3$ $\mu$s and the tail of the wave leaves the cloud at $t=52.5$ $\mu$s. During the passage of the wave, the void fraction oscillates between 0.5$\times10^{-3}$ and 1.0$\times10^{-3}$ due to excitations by the alternate compression and tension in the wave.
After the passage of the wave, the bubbles continue to expand and the void fraction reaches its maximum value: $1.41\times10^{-3}$ at $t = 53.6$ $\mu$s, before decaying to its initial value by $t = 80$ $\mu$s.

\begin{figure}
\center
  \includegraphics[width=125mm,trim = 00mm 0mm
  00mm 0mm,clip]{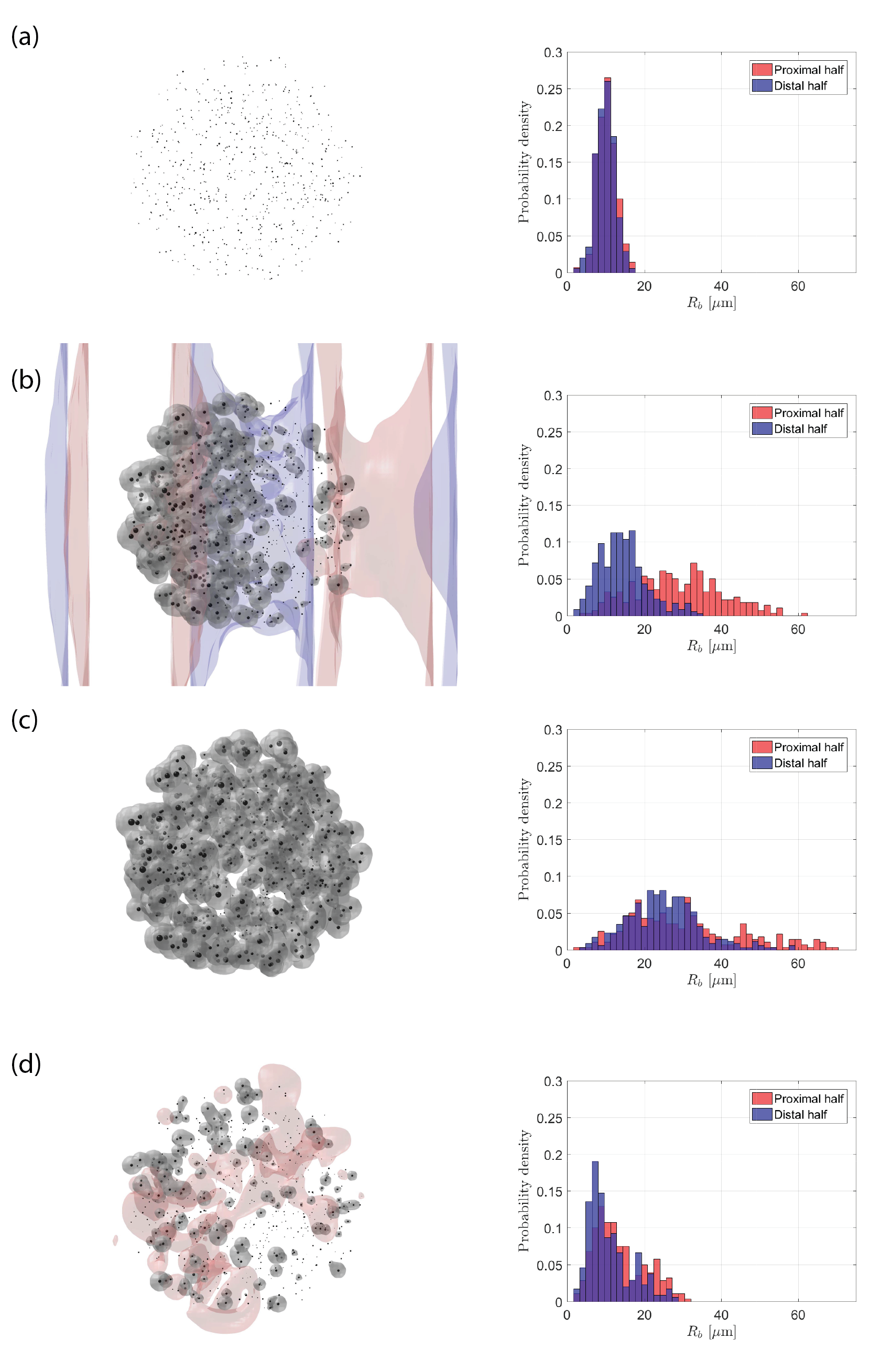}
  \caption{Snapshots of the bubble cloud during the simulation using the three-dimensional model. (a) initial condition, (b) $t=37$
  $\mu$s (during the passage of the wave), (c) $t=53.6$ $\mu$s (timing at the maximum cloud volume) and (d) $t=65$ $\mu$s (after the passage of the wave during the collapse).
  Four quantities are shown:
  the surface of bubbles by black color; iso-contour of void fraction with a value
  of $10^{-3}$ by white color; pressure iso-contours with the values of $-10^5$
  and $10^5$, by blue and red colors. The bar charts at right show the the distribution of bubble size
  in the proximal and distal halves of the cloud at the corresponding times.}
   \label{fig:3dbubble} 
\end{figure}

Fig.\ref{fig:3dbubble} shows the images of the bubble cloud obtained with the three-dimensional model, at various stages during the evolution as well as measured probability distribution functions (PDFs) of the bubble radii in the proximal ($x<0$) and distal ($x>0$) half of the cloud at the corresponding times.
During the passage of the wave, it is clearly seen that the bubbles in the proximal half of the cloud are larger than the bubbles in the distal half.
This is confirmed in the PDF of the bubble radius in that the proximal half presents much larger peak value and broader distribution than that in the distal half.
Meanwhile, around the time of maximum void fraction and subsequent collapse, the PDF is more uniform across the halves. The radial distributions of the bubbles in the two halves are similar to each other. Though the proximal half has larger radii than the distal one.
Fig\ref{fig:3dbubble}d captures the pressure waves generated by the bubbles during the collapse.

\begin{figure}
\center
  \includegraphics[width=140mm,trim = 00mm 0mm
  00mm 0mm,clip]{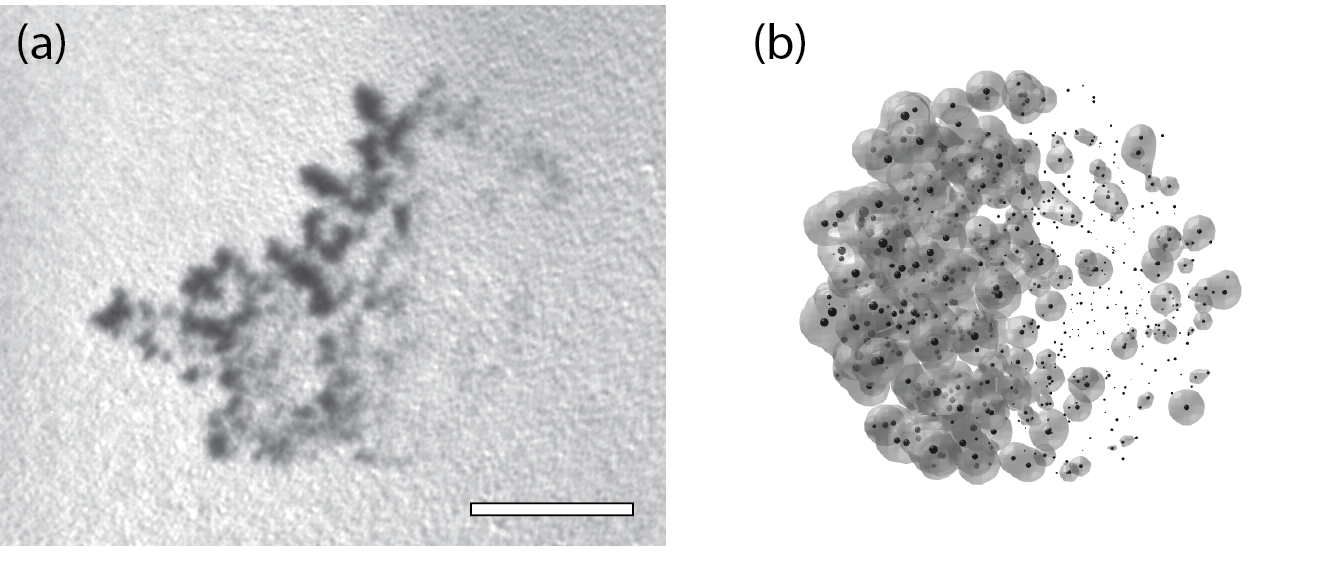}
  \caption{(a) High-speed image and (b) the surface of bubbles and iso-contour of void fraction with a value in the present three-dimensional simulation at $t=37$ $\mu$s. The scale-bar denotes 2 mm.}
   \label{fig:imacon} 
\end{figure}

In fig.\ref{fig:imacon} we are comparing a high-speed image of a bubble cloud excited by a focused ultrasound pulse generated by a medical, piezo-ceramic transducer in water and the snapshot of bubble cloud in the present simulation at $t=37$ $\mu$s.
The pulse travels from the left to the right in the image. 
The pulse contains a train of 10 cycles of a sinusoidal pressure wave with a frequency of 335 kHz and a peak maximum amplitude of 6 MPa. The image was taken at $30$ $\mu$s after the arrival of the wave at the center of the bubble cloud.
In both images, it is shown that the only proximal half of the bubble cloud is excited.

\begin{figure}
\center
 \begin{subfigmatrix}{1}
  \subfigure[]{\includegraphics[width=125mm]{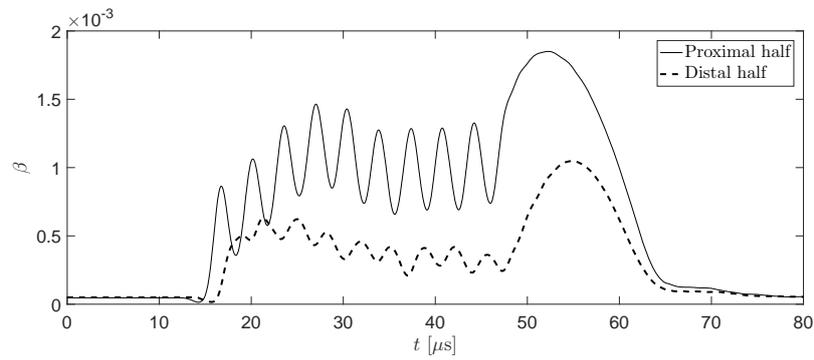}}
  \subfigure[]{\includegraphics[width=125mm]{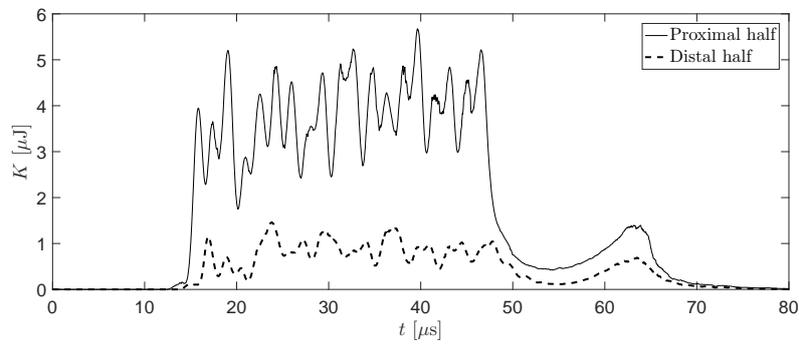}}
  \end{subfigmatrix}
  \caption{Evolution of (a) the void fraction and (b) the kinetic energy of liquid induced by the bubbles in the proximal and distal halves of the cloud.}
   \label{fig:3d_vk} 
\end{figure}

In order to further quantify the anisotropy of the bubble cloud,
in fig. \ref{fig:3d_vk} we plot the evolution of the void fraction and the component of the kinetic energy, induced by the radial oscillations of bubbles, in the proximal and distal halves of the cloud, respectively.
The energy, $K$, is obtained by superposing the component of the kinetic energy in an incompressible flow outside each of the radially oscillating bubbles that reside in the proximal or distal half of the cloud, respectively \citep{Caflisch85}:
\begin{equation}
K=2\pi\rho\sum R_n^3(\vector{x}_n,t)\dot{R}_n^2(\vector{x}_n,t)
\left\{
  \begin{array}{@{}ll@{}}
    x_n<0, & \text{Proximal half}\\
    x_n>0, & \text{Distal half}
  \end{array}\right
  .
\end{equation}

As shown in fig. \ref{fig:3d_vk}a, during the passage of the wave, the void fraction in the proximal half oscillates around a value of 1.0$\times10^{-3}$, while that of the distal half stays around at 0.5$\times10^{-3}$, with a smaller amplitude of the oscillation. The phase of the oscillation in the distal half is delayed from the proximal half, due to the delay in the arrival of the incoming wave.
Both of the halves experience growth and decay after the passage of the wave, yet the growth in the distal half is smaller by approximately 50\%.
During the passage of the wave, the value of the kinetic energy oscillates around at 4.0 $\mu$J in the proximal half, while
it oscillates around at 1.0 $\mu$J in the distal half, with milder oscillations.
The cloud's structure is the result of a shielding of the distal bubbles by the proximal ones.
In other words, the proximal bubbles absorb and scatter acoustic energy such that the incident pulse is attenuated before it interacts with the distal bubbles.
After the passage of the wave, the kinetic energy in the both halves decay to the local minimum at around $t=55$ $\mu$s.
This decay corresponds to the decay in the radial velocity of the bubbles, when the volumetric oscillations of bubbles transit from growth to collapse.
Subsequently, the kinetic energy in the both halves grows to take the local peak at around $t=63$ $\mu$s, then decays back to zero.
This simultaneous peaking corresponds to the coherent collapse of the cloud observed in fig. \ref{fig:3d_vk}.

During the treatment of lithotripsy, the energy shielding of kidney stones caused by bubble clouds may result in a decreased efficacy of stone comminution, and thus is a critical factor for the success of the treatment.
Direct observation of the anisotropy of an acoustic cavitation bubble cloud due to the energy shielding in the numerical simulation has not, to our knowledge, been achieved in previous studies.
The present method can be potentially useful to quantify the energy shielding for applications to lithotripsy and other ultrasound therapies.

\section{Conclusion}
\label{section:conc}
We constructed a coupled Eulerian-Lagrangian method for simulation of cloud cavitation in a compressible liquid.
The mixture-averaged equations are discretized on an Eulerian grid, while individual bubbles are tracked as Lagrangian particles.
The strong, bubble-scattered pressure waves propagating in the continuous phase are accurately captured on the grid by using a WENO-based flow solver, while the radial oscillations of bubbles are evolved by solving the Keller-Miksis equation at the sub-grid scale.
Dimensional reduction of the model was achieved for cases where the bubbly mixture possesses spatial homogeneities, by descritizing the field into two-dimensional or axi-symmetric grids, and modeling the resulting missing bubble-induced pressure fluctuations at the sub-grid scale as white noise.
The method is capable of capturing the multi-scale dynamics of cloud cavitation, including the pressure fluctuations at the scale of single bubble and fine structures of a bubble cloud excited by a strong ultrasound wave.
Such features of the method can be useful in various applications, such as evaluation of the damage potential on materials due to the bubble collapse as well as computations of the effective, total acoustic energy delivered to a target under the presence of cavitation bubbles, during ultrasound therapies.
Future improvements can include non-sphericity and fusion/break-up of bubbles, and further verifying/improving the sub-grid models of the pressure fluctuations in the reduced models for various cases of cloud cavitation and cavitating flows.

\section*{Acknowledgments}
The authors thank Vedran Coralic and Jomela Meng for valuable discussions on the finite volume WENO scheme,
Daniel Fuster for helpful discussions on the bubble-dynamic closure, and Wayne Kreider, Adam Maxwell and Michael Bailey for their support in the companion experiments.
K.M would like to acknowledge the Funai Foundation for Information
Technology, for the Overseas Scholarship.
This work was supported by the National Institutes of Health under grant
2P01-DK043881.
The three-dimensional computations presented here utilized the Extreme Science
and Engineering Discovery Environment, which is supported by the National
Science Foundation grant number CTS120005.

\section*{Citations}
\bibliographystyle{revtex}


\section*{Appendix}
\subsection*{Appendix A. Numerical algorithm}
\label{appd:alg}
In this appendix, we summarize the numerical procedure.
The sequence of steps in pre-processing and simulation using the proposed method is outlined as follows.\\

\noindent1. Pre-processing

(a) Initialize $\vector{q}_{l,i,j,k}$ on a grid given the initial condition.

(b) {\it If two-dimensional or axi-symmetric simulation}: generate random phase $\phi_{k_{i,j}}$.

(c) Initialize Lagrangian variables $R_n$, $\dot{R}_n$, $p_{Bn}$, and $m_n$, given the initial condition.\\
\\
\noindent2. Simulation\\
{\it During each RK-step:}

(a) Compute RHS of equation (\ref{eqn:vae_semi}) using the WENO scheme.

(b) Smear $V_n$ and $\dot{V}_n$ on the grid to obtain $\beta$ and $\dot{\beta}$ using the kernel.

(c) Compute $\vector{g}_{i,j,k}$.

(d) Obtain $p_{cell}$ ($Tp_{cell}$ in case of two-dimensional or axi-symmetric simulation) at the coordinate of each bubble.

(e) {\it If three-dimensional simulation}: compute $p_{\infty}$ at the coordinate of each bubble.

(f) {\it If two-dimensional or axi-symmetric simulation}: compute $p'_{cell}$ at each coordinate of the bubble.

(g) Compute $\dot{R}_n$, $\ddot{R}_n$, $\dot{p}_{Bn}$, and $\dot{m}_{Vn}$.

(h) Update $\vector{q}_{l,i,j,k}$.

(i) Update $R_n$, $\dot{R}_n$, $p_{Bn}$, and $m_{Vn}$.

\subsection*{Appendix B. Speedup with the reduced model}
\label{appd:speedup}
Here we quantify the reduction in the computational cost by using the reduced models from the three-dimensional model.
In a test problem, we solve for the dynamics of $N_P$ bubbles distributed in a domain with a size of $x,y,z\in[-2.5, 2.5]$ mm.
The domain is descritized into $N_G=50^3$ finite volume cells for three-dimensional simulations, and $N_G=50^2$ cells for two-dimensional simulations.
For $N_P/N_G\in[10^{-3},10^{-1}]$, we measure the wall time required to march the governing equations by a single time step by using the three-dimensional model, the two-dimensional model, and the two-dimensional model with $p'_{cell}=0$, namely $T_{3D}$, $T_{2D}$, and $T_{2D_0}$, respectively.
Then we compute the speed-up in the wall time: $T_{3D}/T_{2D}$ and $T_{3D}/T_{2D_0}$.
For all the test problems, we use a single CPU core of Intel Xeon E2670v3 processor.

\begin{figure}
\center
 \includegraphics[width=85mm]{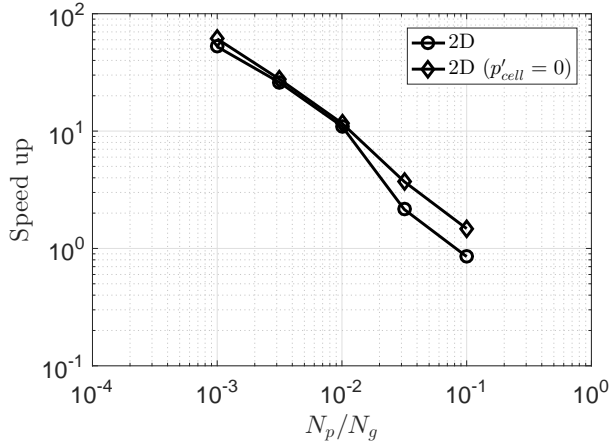}
  \caption{Speed up by using the two dimensional model, with and without modeling $p'_{cell}$.}
  \label{fig:speedup} 
\end{figure}

Fig.\ref{fig:speedup} shows the results.
The speed-up is $O(1-10^{2})$.
In all the cases, the total cost of the simulation is the summation of the cost to compute the Eulerian field and that of the Lagrangian bubbles.
With $p'_{cell}=0$ the speed-up is expected to be globally larger than 1 (the two-dimensional model is always faster), since the cost of time marching the Lagrangian bubbles is the same between the three-dimensional and two-dimensional models except for the smearing procedures, while the cost of the Eulerian field is globally smaller in the two-dimensional model.
By increasing the number of particles, the speed-up decreases, since the cost of the Lagrangian bubbles becomes dominant in the total cost, and the reduction in the cost of the Eulerian phase contributes less to the total cost.
The speed-up with the two-dimensional model with $p'_{cell}$ follows a similar trend as $p'_{cell}=0$.
However, with the same $N_P/N_G$, the speed-up with modeling $p'_{cell}$ is smaller than that with $p'_{cell}=0$.
This difference corresponds to the overhead to compute $p'_{cell}$.
With $N_P/N_G=10^{-1}$, the speed-up is smaller than 1 (the two-dimensional model is slower).
This result indicates that the magnitude of the cost reduction in the Eulerian field is smaller than the overhead of modeling $p'_{cell}$ when $N_P/N_G>O(10^{-1})$, with the particular grid of the test problem.
The result also implies that the two-dimensional model with $p'_{cell}=0$ would be useful in terms of speed-up and cost-reduction, when $p'_{cell}$ does not alter solutions to within an accuracy of interest.

We note that the results also hold in a parallel environment, in which a computational domain, including both the Eulerian field and the Lagrangian particles, is decomposed into sub-domains that are allocated to distinct processors using a Message Passing Interface (MPI) protocol.
We have confirmed a good parallel efficiency of the algorithms.
In that case, the global speed-up is bounded by the speed-up of the sub-domain that has the largest value of $N_P/N_G$.

\subsection*{Appendix C. Details of the sub-grid modeling to obtain $p_\infty$
for the three-dimensional model}
In this appendix, we describe the detailed derivation of the expression of $p_{out}$ (\ref{eqn:p_out}) used in the bubble dynamic closure for the three-dimensional model.
The original sub-grid closure of the Keller-Miksis equation was proposed by FC, in a regime in which multiple-bubbles reside in a single finite volume cell\citep{Fuster11}.
Here we revisit the derivation in a regime where we have at most single bubble in a finite volume cell.
To the aim, we consider a bubble with radius $R_n$ in the control volume $V_{cell}$.
At the sub-grid scale, it can be assumed that the liquid is incompressible and the flow field is irrotational. Thus at an arbitrary coordinate in the liquid in $V_{cell}$, the following Bernoulli's equation holds:
\begin{equation}
\frac{\partial\phi}{\partial t}=\frac{1}{2}(\nabla\phi)^2+\frac{p-p_0}{\rho},
\label{eqn:Bera1}
\end{equation}
where $\phi$ is the velocity potential.
$\phi$ can be decomposed into the velocity potential of the in-coming pressure wave and the out-going wave emitted by the bubble oscillation:
\begin{equation}
\phi=\phi_{\infty}+\phi_n.
\label{eqn:Bera2}
\end{equation}
Note also that, without the presence of the bubble, the Bernoulli's equation can be simplified as
\begin{equation}
\frac{\partial\phi_\infty}{\partial t}=\frac{p_\infty-p_0}{\rho}.
\label{eqn:Bera3}
\end{equation}

By using equations (\ref{eqn:Bera1}-\ref{eqn:Bera3}), we obtain
\begin{equation}
\frac{p-p_\infty}{\rho}=-\frac{\partial\phi_n}{\partial t}+\frac{1}{2}(\nabla\phi_n)^2.
\label{eqn:ber2}
\end{equation}

Though we do not know the value of $p_\infty$ a-priori, it is approximately constant over $V_{cell}$.
The other terms are functions of the distance from the center of the bubble $r$.
In order to eliminate $p_\infty$, we derive two expressions.

First, we write
\begin{equation}
\frac{1}{V_{l,cell}}\int_{V_{cell}}\frac{p-p_\infty}{\rho}dv_l
=
\frac{1}{V_{l,cell}}\int_{V_{cell}}\left[-\frac{\partial\phi_n}{\partial t}+\frac{1}{2}(\nabla\phi_n)^2\right]dv_l,
\end{equation}
where $V_{l,cell}=\int_{V_{cell}}dv_l$.
Notice that
\begin{equation}
\frac{1}{V_{l,cell}}\int_{V_{cell}}p dv_l=p_{cell},
\end{equation}
and thus the relation can be re-written as
\begin{equation}
\frac{p_{cell}-p_\infty}{\rho}
=
-
\frac{1}{V_{l,cell}}\int_{V_{cell}}\frac{\partial\phi_n}{\partial t}dv_l
+
\frac{1}{V_{l,cell}}\int_{V_{cell}}
\frac{1}{2}(\nabla\phi_n)^2 dv_l.
\label{eqn:Brcell}
\end{equation}
In order to explicitly express the right hand side, we approximate the integrals by assuming that the control volume $V_{cell}$ is a sphere with a radius of $R_{cell}=(3/4\pi V_{cell})^{1/3}$ and the bubble resides at the center of the sphere.
By doing so, we can approximate the integral operator:
\begin{equation}
\int_{V_{cell}}(\cdot)dv_l\approx\int^{R_{cell}}_{R_n}(\cdot)4\pi r^2dr.
\end{equation}
We naturally have $V_{l,cell}=\frac{4}{3}\pi (R_{cell}^3-R_n^3)$.
The integrands of the RHS of equation (\ref{eqn:Brcell}) can be expressed in terms of $r$ and the states at the surface of the bubble:
\begin{equation}
\frac{\partial\phi_n(r)}{\partial
t}=\frac{R_n}{r}\frac{\partial\phi_b(R_n)}{\partial t},
\hspace{5pt}
\nabla\phi_b(r)=\frac{R_n^2\dot{R_n}}{r^2}.
\end{equation}
Substituting these expressions, we obtain
\begin{align}
\frac{1}{V_{l,cell}}\int_{V_{cell}}\frac{\partial\phi_n}{\partial
t}dv_l
&\approx
\underbrace{\frac{3}{2}\frac{R_n(R^2_{cell}-R^2_n)}{R^3_{cell}-R^3_n}}_{C_1(R_n,R_{cell})}
\frac{\partial\phi_b(R_n)}{\partial
t},\\
\frac{1}{V_{l,cell}}\int_{V_{cell}}\frac{1}{2}(\nabla\phi_b(R_n))^2 dv_l
&\approx
\underbrace{\frac{3}{4}\frac{R_n^3}{R^3_{cell}-R^3_n}\left(1-\frac{R_n}{R_{cell}}\right)}_{C_2(R_n,R_{cell})}\dot{R}_n^2.
\end{align}
Finally, we can re-write the relation (\ref{eqn:Brcell}) as
\begin{equation}
\frac{p_{cell}-p_\infty}{\rho}
\approx
-
C_1\frac{\partial\phi_n(R_n)}{\partial t}
+
C_2R_n^2.
\label{eqn:cell}
\end{equation}
This equation represents the averaged contribution of the bubble dynamics to the pressure in $V_{cell}$.

The second equation needed to estimate $p_\infty$ is simply equation (\ref{eqn:ber2}) evaluated at the surface of the bubble:
\begin{equation}
\frac{p_n-p_\infty}{\rho}=-\frac{\partial\phi_n(R_n)}{\partial t}+\frac{1}{2}R_n^2,
\label{eqn:bubble}
\end{equation}
where we used $p(R_n)=p_n$ and $\nabla\phi(R_n)=\dot{R_n}$.
This equation represents the contribution of the bubble dynamics to the pressure at the surface of the bubble.

Now that we have two unknown variables, $\partial\phi(R_n)/\partial t$ and $p_\infty$, in two equations.
It is straightforward to eliminate the unknowns to obtain
\begin{equation}
\frac{1}{V_{l,cell}}\int_{V_{cell}}p_{out}dv_l=p_{cell}-p_\infty\approx\frac{1}{1-C_1}\left[p_{cell}-p_n-\left(C_2-\frac{1}{2}\right)\rho \dot{R}_n^2\right].
\end{equation}







\end{document}